\documentclass[twoside,11pt]{article}

\usepackage[abbrvbib]{jmlr2e}

\newcommand\inner[2]{\langle #1, #2 \rangle}
\newcommand\selfinner[1]{\langle #1 \rangle}

\usepackage{booktabs} 
\usepackage{array}
\newcolumntype{L}{>{\centering\arraybackslash}m{12.5cm}}
\newcolumntype{S}{>{\centering\arraybackslash}m{1.5cm}}
\usepackage{xfrac}
\usepackage{bm}
\usepackage{amsmath}
\usepackage{amsfonts}

\usepackage{amssymb}
\usepackage{bbm}
\usepackage{algorithm2e} 
\usepackage{booktabs} 
\SetKwInOut{Parameter}{Parameter}

\DeclareMathOperator*{\argmin}{arg\,min}

\makeatletter
\newcommand\makebig[2]{%
  \@xp\newcommand\@xp*\csname#1\endcsname{\bBigg@{#2}}%
  \@xp\newcommand\@xp*\csname#1l\endcsname{\@xp\mathopen\csname#1\endcsname}%
  \@xp\newcommand\@xp*\csname#1r\endcsname{\@xp\mathclose\csname#1\endcsname}%
}
\makeatother

\makebig{Biggg} {3.0}

\usepackage{bbm}

\usepackage[utf8]{inputenc}
\usepackage[english]{babel}
 
\usepackage{hyperref}
\hypersetup{
    colorlinks=false,
    linkcolor=blue,
    filecolor=magenta,      
    urlcolor=blue,
}
 
\usepackage{hyperref}

\ShortHeadings{What can the millions of random treatments [...] reveal about causes?}{Ribeiro, Neffke and Hausmann}
\firstpageno{1}

\begin{document}

\title{What can the millions of random treatments in nonexperimental data reveal about causes?}

\author{\name Andre F. Ribeiro \email andre\_ribeiro@hks.harvard.edu \\
       \addr  Harvard University \\ 79 John F. Kennedy Street, Cambridge, MA USA 02138
        \AND
       \name Frank Neffke \email frank\_neffke@hks.harvard.edu \\
       \addr  Harvard University \\ 79 John F. Kennedy Street, Cambridge, MA USA 02138
       \AND
       \name Ricardo Hausmann \email ricardo\_hausmann@hks.harvard.edu \\
       \addr  Harvard University \\ 79 John F. Kennedy Street, Cambridge, MA USA 02138}

\editor{-}

\maketitle

\begin{abstract}
We propose a new method to estimate causal effects from nonexperimental data. Each pair of sample units is first associated with a stochastic 'treatment' - differences in factors between units - and an effect - a resultant outcome difference. It is then proposed that all such pairs can be combined to provide more accurate estimates of causal effects in observational data, provided a statistical model connecting combinatorial properties of treatments to the accuracy and unbiasedness of their effects. The article introduces one such model and a Bayesian approach to combine the $O(n^2)$ pairwise observations typically available in nonexperimnetal data. This also leads to an interpretation of nonexperimental datasets as incomplete, or noisy, versions of ideal factorial experimental designs.  

This approach to causal effect estimation has several advantages: (1) it expands the number of observations, converting thousands of individuals into millions of observational treatments; (2) starting with treatments closest to the experimental ideal, it identifies noncausal variables that can be ignored in the future, making estimation easier in each subsequent iteration while departing minimally from experiment-like conditions; (3) it recovers individual causal effects in heterogeneous populations. We evaluate the method in simulations and the National Supported Work (NSW) program, an intensively studied program whose effects are known from randomized field experiments. We demonstrate that the proposed approach recovers causal effects in common NSW samples, as well as in arbitrary subpopulations and an order-of-magnitude larger supersample with the entire national program data, outperforming Statistical, Econometrics and Machine Learning estimators in all cases. As a tool, the approach also allows researchers to represent and visualize possible causes, and heterogeneous subpopulations, in their samples.
\end{abstract}


\section{Introduction}

Most questions of interest in the social, behavioral and life sciences -- What makes economies grow? What explains criminal behavior? What can prevent or cure a disease? -- are ultimately questions about what \emph{causes} an outcome of interest. Data used to answer such questions typically have no shortage of correlational patterns, but correlations are often poor guides to the causal process that produced them. The central methodological difficulty in scientific inquiry remains that of estimating the \emph{causal effect} of a treatment, or independent variable, on an outcome. Compared to the tremendous success of Machine Learning algorithms in prediction and pattern recognition tasks in correlation-rich data, such as in face recognition and textual topic modelling, Machine Learning approaches are still of limited use when estimating causal effects ~\citep{PearlJudea2019Tsto,Athey483}. This has led to a paradoxical situation: in the midst of the big-data revolution, many prominent scientists have declared randomized field experiments - with often just hundreds of participants - as the sole standard for empirical research ~\citep{imbens,duflo}. Experiments are attractive because, once individuals have been randomly divided into treated and nontreated subgroups, it suffices to compare their average outcomes to estimate causal effects. Yet, randomized trials have many drawbacks: they are expensive, slow and sometimes impossible or unethical to carry out and they elucidate only \textit{if} treatments work, rather than \textit{why} they work ~\citep{deaton,imbens,RefWorks:doc:5addefb1e4b0c16216f35bc0}. Furthermore, a focus on average effects creates problems when different individuals experience different effects. Indeed, Xie ~\citep{RefWorks:doc:5ae0f6f3e4b07da0d123c2b5} (2016, p. 6263) considers this a fundamental conundrum: 'the ubiquitous presence of individual-level variability [in social phenomena] makes it impossible to study individual-level causal effects. To draw a causal inference, it is necessary to pool information from different members in a population into aggregates'. As we will demonstrate, effect heterogeneity affects the accuracy of current observational methods even in datasets of moderate size. While discourse about causes are often dominated by all-or-nothing hypothesis testing in the Sciences, there is great practical need for tools that can introduce causal insights into the earlier phases of scientific discovery or provide insights from larger data. Here, we study a problem representation and method that facilitates the use of recent Machine Learning and high-dimensional techniques to that end.    

\subsection{Model Summary}

In particular, we consider the problem of estimating Average Treatment Effects (ATE) or Individual Treatment Effects (ITE) of a given treatment on an outcome, $y \in \mathbb  R$. The problem of estimating the effect of a treatment-of-interest nonexperimentally have been studied extensively, in particular from comparisons between treated and non-treated subjects' outcomes ~\citep{RefWorks:doc:5911ec38e4b0ac17f7d9f452,match-survey,RefWorks:doc:5aa5b99fe4b0db66bcfcdb5c}. Although for nonexperimental estimation, these approaches often draw on Experimental Design concepts and have received attention, especially, in Econometrics and Applied Statistics. 

The central goal of the present work is to better understand and exploit the heterogeneity of statistical conditions between pairs of individuals in everyday datasets, as they relate to causality. Consider a nonexperimental dataset with $n$ observations and $m$ variables, from a variable set $\mathcal{X}^m$. An observed difference in outcome $y_{ij} = y_i - y_j$ between any two individuals, $ 0 < i,j \leq n$, can carry both a lot or very little information about a variable, or variable subset, $v \subset \mathcal{X}^m$. When individuals differ only by a single variable $v=\{a\}$, strict claims can be made for the effect of $a$, as the pairing characterizes an ideal counterfactual (given conditions reviewed below). More commonly, however, pairwise conditions are in a spectrum of usefulness for each variable and individual. Imagine the existence of a prior stochastic process that can describe how individual pairwise conditions translate to the validity of pairwise observations, $y_{ij}$, for each  variable subset $v$. Such model would allow us to use all pairs, and the full range of experiments 'run by nature', to help estimate causal effects.  

An obvious analogy is to Signal Processing and Fusion ~\citep{MacKay:2003aa,Hall:2008aa}, where the use of \textbf{millions of (noisy) observations often outperform any individual observation by orders of magnitude – provided a suitable prior statistical model for measurements}. We characterize all population pairwise conditions by considering all their possible combinations of differences and intersections. The model then connects such combinatorial conditions to the validity of pairwise estimates.  The effect $f(v)$ of variables $v$ is first described as $f(v) \sim \mathcal{N}(y_{ij}, \sigma_{ij}(v))$, which carries the assumption of Gaussian measurements having a common mean, for each $v$, but distinct standard deviations $\sigma_{ij}$ across pairs. In datasets below, there are in the order of 50-200M such pairs. Here, variables $y_{ij}$ and $v$ are assumed observable. We relate the unobservables, $f(v)$ and $\sigma_{ij}(v)$, to the asymmetric differences of pairs under its two commutations – corresponding to the effect of 'treatments' and their observed confounders. We also show these have straight-forward interpretations as vector angles in covariate space. Since each pair is seen as a deviation from an ideal factorial experimental run, it becomes useful to characterize a nonexperimental dataset, overall, as a noisy version of an ideal Factorial Experimental Design. 


\begin{figure}
\centering
\includegraphics[width=0.8\linewidth]{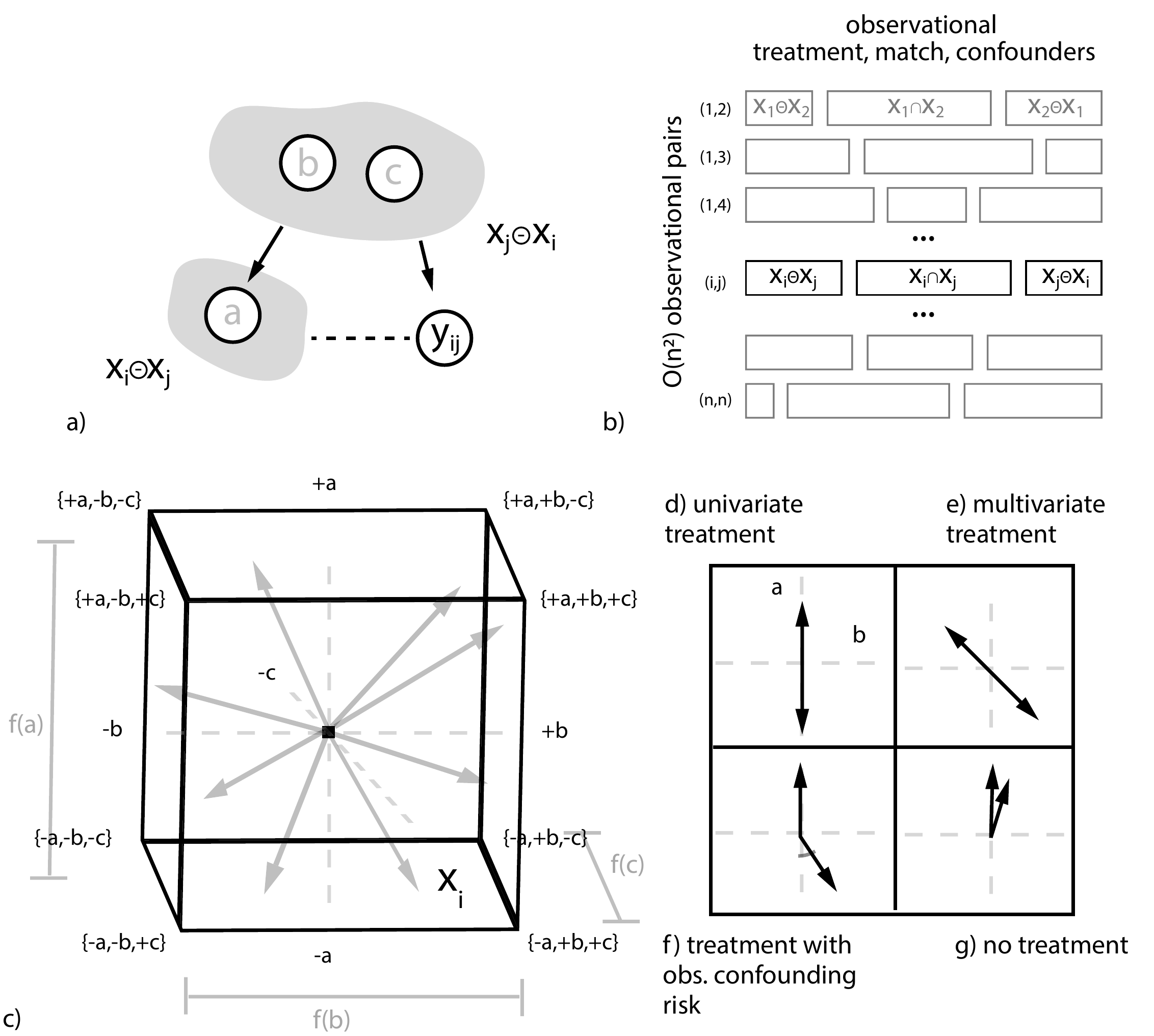}
\caption{\label{fig-fract-intro} \textbf{(a)} for a set of population attributes $\mathcal{X}^3=\{a,b,c\}$, where $x_i \subset \mathcal{X}^3$ is a subset describing individual $i$ (resp. $j$) and $\ominus$ is set difference, illustration for the sample balance condition $x_i{\ominus}x_j \perp y \,\vert\, x_j{\ominus}x_i$ for a pairwise difference $x_i{\ominus}x_j = \{a\}$ ('treatment') between two individuals $(ij)$, with $x_j{\ominus}x_i = \{b,c\}$, \textbf{(b)} for a $n$-sized nonexperimental sample and $m$ factors, decomposition of the sample $\sfrac{n(n-1)}{2}$ pairs into the same number of treatments ($x_i{\ominus}x_j$), matches ($x_i\cap x_j$) and possible confounders ($x_j{\ominus}x_i$), \textbf{(c)} Factorial cube $\mathcal{C}^3$ with factors $\mathcal{X}^3 = \{a,b,c\}$ which can assume high and low (Boolean) values, notated $\texttt{+}a$ and $\texttt{-}a$, $\mathcal{C}$'s vertices correspond to all possible factorial experimental \emph{runs} and edges to \emph{treatments}, vectors (gray) depict $8$ individuals' positions according to a $8 \times 3$ standardized observation matrix $X$, $f(a)$ is the causal effect of variable $a$ (resp. $b,c$) in the proposed representation; \textbf{(d-g)} representative individual pairings and their implied factorial treatments.}
\end{figure}

Using the proposed combinatorial population model, we also derive other sources of measurement 'noise' from the causal effect estimation literature. The way we formulate such added deviations is purposively standard, allowing us to later benchmark the proposed approach (using all pairs) to alternatives selecting pairs or other popular estimators. It would be easy to add more intricate conditions, that can be formulated combinatorically, but we show that simple conditions outperform more complex estimators. This is the case, for example, of estimators designed to account for unobservables and endogeneity, when replicating the outcome of the studied Randomized Experiments. 

A popular recent strategy ~\citep{NIPS2017_7223,Wang:2019aa,Abadie:2015aa} is to stipulate prior models assuming a proxy confounding variable (either observed or unobserved) in the data. Latent-Confounder approaches require assuming and training complex functional models to infer the latent variable, then adjusting effect estimates accordingly. In practice, pairs of individuals vary in how they differ, and consequently in what variables can confound estimates in each case. Instead of requiring a single variable to account for this potentially complex and contingent relationship, a prior statistical model for pairwise conditions allow estimates to be considered in a pair-by-pair case - simply penalizing pairs, or observations $y_{ij}$, that are ruled unfit in some fundamental way.



This article's main contribution is therefore to demonstrate the practical benefits of modeling and combining large numbers of disparate pairwise observations. Beyond practical usefulness, understanding how population combinatorial patterns relate to causality is also of great importance to population-based research, such as the study of genetic variation in the Medical Sciences and demographics in the Social Sciences. We discuss implications of concepts developed here to the study of genetic population variation, and Genome-wide Association studies, in ~\citep{ribeiro-genetics-ev}.

\subsection{Sample Balance and Treatment-Assignment Ignorability Conditions}

The previous discussion leads to a simple but also very conservative model, where the risk of (observed) confounding increases with extraneous variables between pairs. We also extend the model to account for common, more specific, conditions from the literature, such as reduced noise for pairs satisfying treatment-covariate sample independence conditions and with small treatments. Let $x_i \subset \mathcal{X}^m$ be a set of attributes describing an individual $i$ (resp. $j$) and $\ominus$ indicate (asymmetric) set-difference.  The former condition, $x_i{\ominus}x_j \perp y \,\vert\, x_j{\ominus}x_i$, is illustrated in Fig.\ref{fig-fract-intro}(a). In the previous model, variables $x_j{\ominus}x_i$ were assumed capable of confounding pairwise estimates, affecting both the treatment and outcome at the same time. With this added assumption, the pairs where treatment-outcome variables are independent, conditional on non-treated variation, are now also deemed useful. Each such assumption thus further scales the number of useful pairs. This condition is, abstractly, 'the most developed and popular strategy for causal analysis in observational studies' ~\citep{Pearl:2010aa,King:2019aa}. It is used by a wide range of methods across disciplines. It requires, however, a further 'ignorable treatment assignment' assumption ~\citep{RefWorks:doc:5a514c24e4b0eeb35a49d1bb,instrumental,NIPS2017_7223}. Namely, it requires that, conditional on the observed variables, there are no unobserved differences between treatment and control groups. The common way to satisfy this assumption is to include in $\mathcal{X}^m$ any variable that affect either outcomes or treatments. This is because, theoretically at least, for any variable set satisfying  ignorability, any superset will too ~\citep{VanderWeele:2011aa}. Due to the centrality of this condition in the causal effect estimation literature, and its further required assumptions, we briefly review them.  

In randomized experiments, randomization enables unbiased estimation of treatment effects across local population groups. For each observed variable, randomization implies, as a simple application of the law of large numbers, that any treatment-subgroups will be, what is often called, 'balanced' on average. Unfortunately, the assignment of treatments to subjects is typically not random in observational datasets. Most causal effect estimator in use today attempt to reduce the treatment assignment bias, and mimic randomization, by increasing a balance score between treatment and control units in use. The idea is to create a subsample of sample units that received the treatment that is comparable on all observed covariates to units that did not receive the treatment.

A balancing score $b(X)$ is a function of the observed covariates $X$ such that the conditional distribution of $X$ given $b(X)$ is the same for treated and control units - thus reflecting the previous assumption. There are three general approaches to derive such functions. The first and most trivial function is $b(X)=X$, which is the case of exact matching (we review these methods in detail below). A second approach, which broadly underlie popular Propensity Score and Latent-Confounder estimators, is to use dimensionality reduction techniques to define a simpler function $\pi(X)$. With this function, it is easier to balance samples, $[ v \perp y \, |\, X ] \approx [ v \perp y \, |\, \pi(X)]$. On the other hand, the approach requires not only the added assumption, but also more complex models, both statistically and computationally. With ignorability, it has been shown analytically ~\citep{Rosenbaum:1983aa} that Propensity scores are the 'coarsest' balancing function taking the multidimensional $X$ into one dimension. It uses a logistic regression to calculate the probability of a unit being assigned to a particular treatment, given $X$. A third approach is to use explicit, or non-parametric, balance functions. The simplest such function is a difference-of-means (i.e., between covariate means across treated and nontreated samples). There are additional analytic advantages to this approach ~\citep{King:2019aa} - such as, concurrently, decreasing model dependence. Due to the focus on large numbers of pairwise observations, we favor this alternative. That is, for a large number of pairs, it is undesirable to run a large number of regressions\footnote{We would also mention that we experimented with other alternatives, such as HSIC tests and Distance Correlations ~\citep{RefWorks:doc:5ad03a32e4b04fa36b7310cf}, while results with these alternatives are comparable to those used in reported results, we favor explicit balance scores due to their computational efficiency and very natural interpretation in the geometrical framework formulated below.}. 


The assumption of ignorability that often accompanies observational methods has been challenged recently, such as when not all relevant variables can be included ~\citep{Wang:2019aa,Athey:2019aa,NIPS2017_7223}. \textbf{While sometimes calling confounders 'unobserved', these parametric estimators assume partially-observed confounding variables} (whose correlations with observed variables can be exploited when training a proxy). Beyond analytical discussions, there is however a more practical problem underlying ignorability. Popular ATE estimation methods such as Propensity Scores are often sensitive to inclusion of non-causal variables ~\citep{RefWorks:doc:5a514530e4b0e3e5a635f08a,RefWorks:doc:5a5148e4e4b0e3e5a635f0e7,RefWorks:doc:5a515a0ee4b0e3e5a635f2cc}. These opposing constraints (to include as many variables as possible versus not including non-causes) create practical difficulties for researchers and threaten the validity of observational estimates - especially in datasets with many variables. Addressing selection bias is essential to identify extraneous effects of other causes on the treatment-of-interest, but does not address the second problem. Conducting model selection and effect estimation in a common framework is a promising direction to rule out non-causes ~\citep{ChernozhukovVictor2015PaPI}. Because of the use of pairwise outcome differences, $y_{ij}$, we argue it is natural to translate these two problems into distance metric learning problems. The approach leads, as a result, to a data representation that is more easily interpretable by researchers, reflecting Factorial Experimental Designs. Other immediate advantages of the approach are discussed below.


\subsection{Reproducing Effect Estimates from Large Randomized Experiments}


We assess the proposed method's performance in simulations and a seminal real-world example, comparing it to current Statistics, Econometrics and Machine Learning estimators. We demonstrate that the proposed approach also remain accurate in heterogeneous and diverse samples. The simulations introduce confounders and heterogeneous subpopulations into synthetic data, demonstrating that observational methods generally become biased or inaccurate, unlike the proposed. As real-world application, we consider the National Supported Work (NSW) program. Starting with a seminal contribution by Lalonde ~\citep{RefWorks:doc:5a5144d2e4b08e15c00ca6cf}, studies have used this Randomized Control Trial to benchmark nonexperimental techniques - including an historical 'face-off between regression and propensity-score matching' ~\citep{harmless}. Observational methods generally fail to recover the experimental causal effect estimate, except in a smaller handpicked NSW subsample ~\citep{RefWorks:doc:5a514530e4b0e3e5a635f08a,RefWorks:doc:5a5145b6e4b01d3dd55629f7,RefWorks:doc:5a514994e4b08e15c00ca73d,RefWorks:doc:5aa5b99fe4b0db66bcfcdb5c}. This literature exemplify a typical scenario across disciplines: estimating causal effects nonexperimentally require several (hard to justify) population and variable selection assumptions (in this case, expert selection of samples and variables with desirable economic characteristics). We demonstrate, however, that the proposed approach can recover the NSW experimental effects not only in Lalonde's original unsolved challenge (with 740 participants and 6 variables) but also in the full NSW data with over 10000 participants and 1000 variables, without ex ante assumptions from researchers. 

We compare the approach to a range of previous solutions, including those making typical ignorability assumptions, as well as approaches relaxing other assumptions, such as missing variables and endogeneity. We show that the previous simple model outperform these solutions, when trying to reproduce the results of the previous Randomized Experiment. Our initial goal was to address the common case of large datasets with many variables. Perhaps surprisingly, however, this is also the case for samples with very few variables, where Instrument-based and Latent-Confounder approaches should be most relevant.

\section{Stochastic Factorial Estimation (SFE)}\label{sect-sfe}

\textbf{We first introduce the proposed approach, then review the related literature in further detail.}
 Randomized Controlled Trials (RCTs), and the observational estimators they have inspired ~\citep{RefWorks:doc:5a514c24e4b0eeb35a49d1bb}, often focus on the causal effect of a single treatment or intervention. However, observed outcomes are often the result of many interacting causes. This limitation of RCTs had already been noted by Fisher in 1926: 'No aphorism is more frequently repeated in connection with field trials, than that we must ask Nature few questions, or, ideally, one question, at a time.' ~\citep{fisher} (1926, p. 503) Instead, he proposed submitting Nature 'logical and carefully thought out questionnaire[s]', leading to \emph{factorial experimental designs}. Factorial designs have since been mostly studied for the design or analysis of experiments ~\citep{DasguptaTirthankar2015Cif2}. 

 A factorial experiment is a complex experiment consisting of many runs. It is designed to estimate the causal effect of $m$ factors on an outcome of interest $y$.  When factors are binary, the design contains $2^m$ \emph{factorial runs}, or, possible factor combinations. Fig.\ref{fig-fract-intro}(c) depicts geometrically a 3-factor design with a cube $\mathcal{C} : \{\texttt{-}1,\texttt{+}1\}^3$. We call the set of factors in which two runs differ a \emph{factorial treatment}. A factorial run corresponds to $\mathcal{C}$'s vertices and treatments to edges. We consider all individuals in a nonexperimental dataset as stochastic factorial runs and the entire dataset as an incomplete random factorial design. Critically, the full set of observed factorial treatments express necessary combinatorial patterns of variable variation and fixation necessary to make claims about each variable's piecewise effects on $y$. 
 
 
 

 We consider this geometric representation for the nonexperimental causal effect estimation problem next, then discuss a Bayesian procedure that combine effect estimates from each pair of individuals, given how strongly they depart from ideal factorial treatments.  We call the resulting method Stochastic Factorial Estimation (SFE).

\subsection{Geometric Representation}

Consider a Factorial Experiment studying the effects of a set of factors $\mathcal{X}^m = \{a,b,c,...\}$. The experiment's runs are the vertices of the cube $\mathcal{C(X)}^m :  \{\texttt{-}1,\texttt{+}1\}^m$. Factorial treatments are pairs of runs that differ on a set of factors (the 'treatment'), while having all other factors in common. Namely, let $x_i,x_j \in \mathcal{C}^m$ be runs and their corresponding treatment be the set of factors run $i$ has exclusively, $x_i{\ominus}x_j$. In addition, we let the size of $\mathcal{C}$'s edges correspond to the causal effect, $f(x_i{\ominus}x_j)$, of their associated treatment. We discuss an extension to the continuous variables case in Appendix B. 


Consider now an observational matrix $X$ with variables $\mathcal{X}^m$. Fig.\ref{fig-fract-intro}(c) depicts an example with $n{=}8$ individuals and $m{=}3$ variables, where values have been normalized to the unit interval, $X : [\texttt{-}1,\texttt{+}1]^m$. The $O(n^2)$ pairs of individuals are in a myriad of configurations. As a result, different pairs are useful for estimating effects of different variables. Fig.\ref{fig-fract-intro}(d) illustrates a pair corresponding to a factorial treatment with $\{a\}$ as treatment. The pair captures the main intuition behind factorial designs: a single variable, $a$, differs between individuals while all other variables are fixed. Fig.\ref{fig-fract-intro}(e) depicts another factorial treatment. Here, however, the treatment is multivariate, $\{ a,b \}$. Because the treatment consists of two potential causes, it is impossible to infer their separate effects \textit{from the pair alone}. However, we can still learn about their combined effect. Fig.\ref{fig-fract-intro}(f) shows an imperfect factorial treatment. There, the outcome difference is not necessarily due to variables in the treatment and may reflect extraneous variation from other variables. Although not fixed within the pair, these nontreated variables could coincide in expectation across treated and non-treated individuals - i.e., they could be 'balanced' in the sample. We can also learn from pairs in this case. Finally, Fig.\ref{fig-fract-intro}(g) illustrates a pair without treatment. We disregard such cases in the estimation. 

We will use observed factorial treatments like the ones depicted in Fig.\ref{fig-fract-intro}(d-g) to iteratively transform $X$, such that distances between individuals come to represent expected treatment effects on $y$,

\begin{equation}\label{eq-goal}
T_y(X^m) =  \Biggg\{\
\arraycolsep=1.4pt\def\arraystretch{2.2}\begin{array}{cc}  
\mathbf{x}_i  \in \mathbb  R^{m}:&\begin{aligned}
&\vert \mathbf{x}_i -\mathbf{x}_j \vert^2 = p_y(x_i,x_j)f(x_i{\ominus}x_j),  &\forall \,\, 0<i,j \leq n. 
\end{aligned} 
\end{array}
\Biggg\},
\end{equation} 

where $T_y(X)$ is a map $[\texttt{-}1,\texttt{+}1]^m \rightarrow  \mathbb{R}^m$, $\mathbf{x}_i$ (in bold) is the transformed position for individual $i$, $p_y(x_i,x_j)$ is the probability that the pairing of $i$ and $j$ reproduces the factorial treatment $v {=} x_i{\ominus}x_j$, and, $f(v)$ is the treatment's effect.  The use of factorial treatments leads to an estimator for outcome differences $y_{ij}$ as distances in $T_y(X^m)$. We discuss $p_y(x_i,x_j)$ and the resulting stochastic model next. 

\subsection{Stochastic Model and Assumptions}\label{sect-stochastic}
 
 Let $0<i,j \leq n$ be any two individuals and $g(x),h(x)$ be two functionals over $x$. We do not postulate a model that relates an individual's characteristics to her outcomes, $y_i = g({x}_i) + \epsilon$. Instead, we interpret $y_{ij} = y_i - y_j$ as a noisy observation of the true causal effect of observed factor differences, $x_i{\ominus}x_j$. Noise increases as $x_i{\ominus}x_j$ departs from factorial, balanced and univariate treatments. Such departure is described by a distribution $p_y(x_i,x_j)$. That is, we interpret differences in characteristics between a pair of individuals, ${x}_i{\ominus}{x}_j$, as `treatment' differences that cause differences in outcomes $y_{ij} =y_i{-}y_j$, 

\begin{equation}\label{eq-main}
\begin{split}  
y_{ij} &\sim g({x}_i {\ominus} {x}_j) + \epsilon_{ij},\\[1.5ex]
  \epsilon_{ij} &\sim\mathcal{N}(0,  h( {x}_j {\ominus} {x}_i) ),
\end{split}  
\end{equation}

where $\epsilon_{ij}$ is a Gaussian noise with mean $0$ and variance $\sigma^{2} = h({x}_j {\ominus} {x}_i)$. Eq. (\ref{eq-main}) postulates that $\epsilon_{ij}$ reflects distortions in observed effects $y_{ij}$ due to variables that individual $j$, alone, has. We will say that, when $\epsilon_{ij}=0$, pairwise observations correspond to factorial treatments: pairwise observations with little risk of observed confounding. Or, similarly, that each observational pair, ${x}_i{\ominus}{x}_j$, in the sample represents a factorial treatment, $v{=}{x}_i{\ominus}{x}_j$, with probability $p_y({x}_i,{x}_j )$. 


A first way to estimate effects $f(v)$ is to focus on pairs that approximate a given factorial treatment $v$ with near certainty: $p_y(v) \rightarrow 1$. The defining characteristic for this type of pair is that, when estimating the effect for an individual $i$, the other individual $j$ has no observed extraneous factors that could confound the effect $y_{ij}$, $x_j{\ominus}x_i = \emptyset$. This is the first key condition behind Factorial experimentation. In this first condition, the experimenter keeps all relevant conditions fixed, except for a \emph{treatment}. Since factors can, however, be beyond the experimenter's control, a second condition is popular: randomize treatment assignment such that subpopulations are balanced in expectation in the treated and nontreated subsamples. Nonexperimental methods using balancing scores are often seen as attempting to reproduce randomized conditions from nonexperimental data ~\citep{Rosenbaum:1983aa,King:2019aa}. 

Randomization is essentially a mechanism to address selection bias. For a treatment indicator $d \in \{-1,+1\}$, selection biases appear when the treatment $d$ is not independent either of other factors, $X$, or the outcome, $y$. In Econometrics ~\citep{heckman-bias}, when $p(d|X,y) = p(d)$ it is said, under ignorability, that the sample is not subject to selection bias. Computational approaches sometimes make a distinction between covariate, $p(d|X) = p(d)$, and outcome, $p(d|y) = p(d)$, induced bias. These distinctions are discussed in detail in ~\citep{cs-bias,Fan:2007aa}. In Propensity scores or Latent-Confounders based approaches, treated and untreated individuals with the same $\pi(X)$ are expected to have similar distributions across \emph{any} observed baseline covariates - reproducing a randomization experimental procedure. This is often used as diagnostic for their outputs. The former approach is mature and has been studied extensively, both theoretically and practically. There are serious questions as to whether the previous goal can typically be achieved in practice, and whether these methods' assumptions, in fact, hold ~\citep{King:2019aa,RefWorks:doc:5a514530e4b0e3e5a635f08a}. Instead of estimating a parametric model $\pi(X)$ for a given treatment, we calculate explicit balance scores for all $O(n^2)$ pairwise treatments (univariate or multivariate), with simple matrix operations. The calculation is repeated thousands of times in a Bayesian optimization procedure that progressively estimate treatment effects. The approach thus uses 'weak' but numerous balancing scores. This reflects its alternative Signal Processing perspective, as opposed to the more typical, based on Model Inference. In the present framework, the notion of sample balance thus leads to an observational pair's \emph{balance}, $\phi^{bl}_{ij} = p(x_i{\ominus}x_j) - p(x_i{\ominus}x_j|x_j{\ominus}x_i)$. We let $p[\phi^{bl}_{ij}=0]$ denote the probability that the observational pair $(ij)$ is not subject to covariate induced bias, in which case it can also be used to estimate effects. This is given a simple geometrical interpretation below.


The implicit goal of causal effect estimation is to devise effect estimates with high external validity. It is worth considering the impact of multivariate treatments on external validity. Multivariate treatment effects estimate the simultaneous effects of all variables in the treatment, $x_i{\ominus}x_j$. Effects need not generalize to the $2^{\vert x_i{\ominus}x_j\vert}$ different instantiations of the treated variables. Under multivariate treatment conditions, it is impossible to attribute effects to any single cause. As a consequence, the cardinality of a treatment, $\vert x_i{\ominus}x_j \vert$, is inversely related to the external validity of the derived effect estimates. The notion leads to the pairs' \emph{treatment size}, $\phi^{cx}_{ij} = \vert x_i{\ominus}x_j \vert$. We let $p[ \phi^{cx}_{ij} = 1]$ denote the probability that the the observational pair $(ij)$ has a univariate treatment, indicating the propensity for higher external validity.

\subsection{Optimization}
We consider that $p_y({x}_i,{x}_j ) \rightarrow 1$ when conditions $p[\phi^{cx}_{ij} = 1]$ and $p[\phi^{bl}_{ij}=0]$ are fulfilled by stipulating Bayesian priors for: sizes of factorial treatments, $x_i{\ominus}x_j$, and balance of non-factorial variations, $x_j{\ominus}x_i$. This Bayesian formulation leads to an objective function $\Gamma({x}_i)$ over individual positions $x_i$ that we later minimize. The overall procedure transforms pairwise distances $\vert x_i{-}x_j \vert$ in $X$ to reflect observed outcome differences, $y_{ij}$, according to the smallest and most balanced factorial treatments. Appendix A contains a detailed derivation of $\Gamma({x}_i)$ from eq.(\ref{eq-main}), similar to those underlying LASSO and Ridge-Regressions ~\citep{ml-book}. The objective has the form  

 \begin{equation}\label{eq-sol-main}
\begin{split}
\Gamma({x}_i) &=  \min_{{x}_i} \sum_{j=1}^n (1+{\phi}_{ij}^{cx})[ \inner{{x}_i}{{x}_j} + \vert y_{ij} \vert]^2 + {\phi}_{ij}^{bl} \inner{{x}_i}{{x}_j}^2,\\
 &\hat{\phi}_{ij}^{cx} = \frac{\vert {x}_i + {x}_j \vert}{m}, \\ 
 &\hat{\phi}_{ij}^{bl} = \vert \frac{1}{n}\sum_{k=0}^n \inner{{x}_k}{{x}_i+{x}_j} \vert,\\
\end{split}
 \end{equation}

where $\inner{.}{.}$ is the dot-product, ${\phi}_{ij}^{cx}$ and ${\phi}_{ij}^{bl}$ are treatment size and balance estimates. When both ${\phi}_{ij}$ terms are zero and $\inner{{x}_i}{{x}_j}$ is negative, the pair corresponds to a factorial treatment, Fig.\ref{fig-fract-intro}(d-g), and the residual $\inner{{x}_i}{{x}_j} + \vert y_{ij} \vert$ is minimized. Term ${\phi}_{ij}^{bl}$ penalizes unbalanced non-factorial treatments, Fig.\ref{fig-fract-intro}(f), and the consequent risk of confounded estimates (given the discussed assumptions). Term ${\phi}_{ij}^{cx}$ penalizes multivariate treatments, Fig.\ref{fig-fract-intro}(e), and the consequent risk of low external validity. The estimators used in the Experimental section, $\hat{\phi}_{ij}^{cx}$ and $\hat{\phi}_{ij}^{bl}$, are simple treatment size and balance estimators derived directly from the previous geometrical representation (Appendix A).  
 
In the output space, $T_y(X)$, the ATE of any factor $a$ can be calculated simply as the difference in coordinates $a$ between the mean position of all individuals with factor $a$, $\bar{\mathbf{x}}^{\texttt{+}a}$, and those without, $\bar{\mathbf{x}}^{\texttt{-}a}$, 

 \begin{equation}\label{eq-ate}
\begin{split}
  \textit{ATE}(a) =& \sqrt{\vert\bar{\mathbf{x}}^{\texttt{+}a} - \bar{\mathbf{x}}^{\texttt{-}a}\vert(a)},   
\end{split}
 \end{equation}
 
for any $a$ and where $\bar{\mathbf{x}} \in T_y(X)$. That is, in $T_y(X)$ coordinate differences correspond to treatment effects and distances to outcome differences (the squared-sum of treatment effects). Since each factor $a$ with non-zero effects divides a population in two subpopulations, $(\texttt{+}a, \texttt{-}a)$, the method's output also gives researchers means to represent and visualize relevant subpopulations in their samples.
    
\subsection{Related Work}\label{sect-related}


In the Sciences, problems of integrating noisy diagnostic measurements are sometimes  called ‘Inverse problems’. Tikhonov regularization approaches, such as Ridge and LASSO regressions, are popular solutions to inverse problems. They perform both variable selection and regularization in order to enhance the prediction accuracy and interpretability of the statistical models they put forward to explain observations. Regularized inverse problems can be seen as special cases of Bayesian inference ~\citep{Tarantola:2005aa}. We devise a solution on this framework, using well-known Bayesian interpretations of the previous solutions. 

Different disciplines can differ in how they approach causality, with the two most popular frameworks ~\citep{RefWorks:doc:5911ec38e4b0ac17f7d9f452} identified as the Pearl ~\citep{RefWorks:doc:5911ec76e4b0ac17f7d9f458} and Rubin ~\citep{RefWorks:doc:5a514c24e4b0eeb35a49d1bb} frameworks. Due to focus on Experimental Design concepts and pairwise comparisons, we review Rubin's framework, also known as the counterfactual (or potential outcomes) formulation. We do not wish to disregard, however, the critical contributions of other frameworks, and the solutions developed under them. 

Consider a treatment indicator variable $d \in \{\texttt{-}1,\texttt{+}1\}$ and that a treated individual $i$ has observed outcome $y_{i}^{\texttt{+}d} \in \mathbb{R}$. The individual treatment effect of $d$, $\textit{ITE}_i(d)$, is defined as the counterfactual outcome difference,

\begin{equation}
 \textit{ITE}_i(d) = y_{i}^{\texttt{+}d} {-} y_i^{\texttt{-}d}.  
\end{equation}

According to the counterfactual framework, it is impossible to observe the outcome that individual $i$ would have had in the counterfactual situation where she would not have been treated, $y_{i}^{\texttt{-}d}$. A Randomized Controlled Trial (RCT) solves this problem with the help of an homogeneity assumption: since the treatment is administered at random, the nontreated subpopulation's outcome serves, in expectation, as a counterfactual outcome for the treated subpopulation. The manipulation allows researchers to calculate average treatment effects easily,

\begin{equation}
  \textit{ATE}(d) = E[y^{\texttt{+}d}] {-} E[y^{\texttt{-}d}].  
\end{equation}

Nonexperimental approaches, in contrast, often require researchers to specify a data generating process (DGP) for the observed data. A DGP specifies at least:

\begin{itemize}
\item[-] a causal model describing how the treatment affects the outcome variable, as well as how other potential causes may confound the treatment's effect;
\item[-] a population in which this model holds.
\end{itemize}


This approach puts, however, 'the cart before the horse': most research is undertaken because the DGP is poorly understood. We address this problem by combining insights from nonexperimental causal effect estimation and model selection.

Experimental Designs (ED) prescribe conditions, to be verified, or manipulated, by researchers, under which outcome differences, $y_{ij}{=}y_i{-}y_j$, are true causal effects.  We considered two commonplace ED conditions. In the first, the experimenter keeps all relevant conditions fixed, except for a \emph{treatment}. This is the distinguishing strategy of Factorial experimentation. In the second, researchers try to maintain population representativeness in their samples. This is the central strategy underlying Randomized experimentation. These two conditions inspired the development of exact ~\citep{match-exact} and balance-based matching estimators ~\citep{heckman-bias,RefWorks:doc:5a514c24e4b0eeb35a49d1bb,genetic} in observational analysis,   

\begin{equation}
\begin{split}
  \textit{ATE}(d) &= E[y_i {-} y_j], \; \mathrm{iff}  \; x_i{\ominus}x_j {=} \{d\}  \;\;\;\;\;\;\;\;\; \mathrm{(exact)}\\
 \textit{ATE}(d) &= E[y_i {-} y_j], \; \mathrm{iff}  \; x_i{\ominus}x_j {=} \{d\}  \; \text{\&}\;   y \perp \{d\} \; | \; \mathcal{X}{-}{d}  \;\;\;\;\;\; \mathrm{(balance{-}based)}
\end{split}
\end{equation}

These estimators try to find pairs fulfilling the previous conditions for a treatment-of-interest, typically, a variable ${d}$. They have been used across disciplines ~\citep{match-survey,RefWorks:doc:5aa5b99fe4b0db66bcfcdb5c} such as statistics ~\citep{rubin2}, epidemiology ~\citep{match3}, sociology ~\citep{match2}, economics ~\citep{RefWorks:doc:5a514ba1e4b0eeb35a49d190} and political
science ~\citep{match6}. 



We defined a combinatorial-based model for random treatments, eq.(\ref{eq-main}), which can be used to combine large numbers of pairwise observations. Using pairwise comparisons, SFE is related to matching estimators ~\citep{RefWorks:doc:5a514c24e4b0eeb35a49d1bb,match-survey} but different in two important ways. Factorial treatments, eq.(\ref{eq-main}), are more specific statistical entities than matches. The approach articulates statistical roles for all possible combinatorial conditions appearing in pairings, such as when there are extraneous varying variables, ${x}_j {\ominus} {x}_i$, as well as multivariate treatments, ${x}_i {\ominus} {x}_j$. The formulation of a random treatment model enables the use of all available treatments and treatment types (such as multivariate treatments) in datasets. To that end, we decomposed all sample pairs into three components: $x_i{\ominus}x_j$ (treatment), $x_i\cap x_j$ (match), $x_j{\ominus}x_i$ (possible confounders). Fig.\ref{fig-fract-intro}(b) illustrates the decomposition. The decomposition leads to a simple geometrical interpretation of observational pairs and a density $p_y(x_i,x_j)$ that indicates pairs' departure from ideal factorial treatments: univariate treatments with no observed confounders (Appendix A discusses this geometric connection in further detail, as well as density versions for categorical and continuous variables). 



This leads to a second important difference. The decomposition allows the method to autonomously reduce the input data dimensionality. A Bayesian procedure first estimates the \emph{effects} of treatments $x_i{\ominus}x_j$ with largest $p_y(x_i,x_j)$, which simplifies the estimation of effects for the remaining treatments, and repeats. Estimation is progressively simplified because variables without any significant effect on $y$ can be ignored in each step, relaxing the conditions-to-be-fulfilled for the remaining variables (such as the penalties, $\phi_{ij}$, for sample balance and treatment sizes). This is an effective strategy because, contrary to matching assumptions, treatments in nonexperimental data are overwhelmingly multivariate or have negligible effects on outcomes. While not directly about the treatment-of-interest, each observational treatment provides some information about which variables have any significant effect on $y$ and which do not. In the Experimental section, we consider both traditional and iterative extensions to balance-based matching methods ~\citep{RefWorks:doc:5aa5b99fe4b0db66bcfcdb5c,RefWorks:doc:5aa6f0c4e4b059d4ac147e0e,genetic}, as well as other recent Econometrics ~\citep{ChernozhukovVictor2015PaPI} and Machine Learning methods ~\citep{NIPS2017_7223,RefWorks:doc:5aa6f0c4e4b059d4ac147e0e}.  

Whereas matching estimators typically focus on the causal effect of a single variable, this considers that, somewhat counter-intuitively, the problem of estimating the effect of \emph{one} causal variable can become easier once we try to estimate the \emph{effect} of \emph{all} causes. This is due to the curse of dimensionality from which matching estimators suffer. Initially estimating the effects only of (sets of) variables that are closest to experimental conditions, allows us to ignore variables estimated as having negligible effects in future iterations, progressively lowering the dimensionality of the matching process. The output of this process, $T_y(X)$, represents effects analogously to factorial designs. The previous model, eq.(\ref{eq-main}), and objective, eq.(\ref{eq-sol-main}), combine \textbf{effect estimation and model selection within the standard framework of sparcity-based model selection} (e.g., LASSO, Unbiased-LASSO, Ridge regressions) (Appendix A).   

Notice that because differences $y_{ij}$ are estimated (as opposed to a functional for $y$ conditional on $X$), eq. (\ref{eq-main}), the approach does not require i.i.d. assumptions on subjects. The assumption of i.i.d. observations, although common in Machine Learning, can bias estimates for models that stipulate a functional relationship between factors and output, $y \sim g(X)$. This is because, if any two groups are misrepresented in a sample, average outcome differences (i..e., effects) will consequently reflect such biases in selection. Optimization is, instead, used here to combine pairwise observations under a simple Bayesian rationale and learn distances, eq. (\ref{eq-goal}). This is a central advantage of pairwise observational estimators ~\citep{match-survey}. The approach can, on the other hand, be potentially sensitive to unobserved pairwise differences, leading, in the present interpretation, to incomplete factorial designs. This is however ameliorated by learning from multi-variate treatments, which are overwhelmingly common in everyday datasets. We consider next how practical the presented approach can be in common conditions, compared to pairwise and non-pairwise solutions.  


\begin{figure}
\centering
\includegraphics[width=1\linewidth]{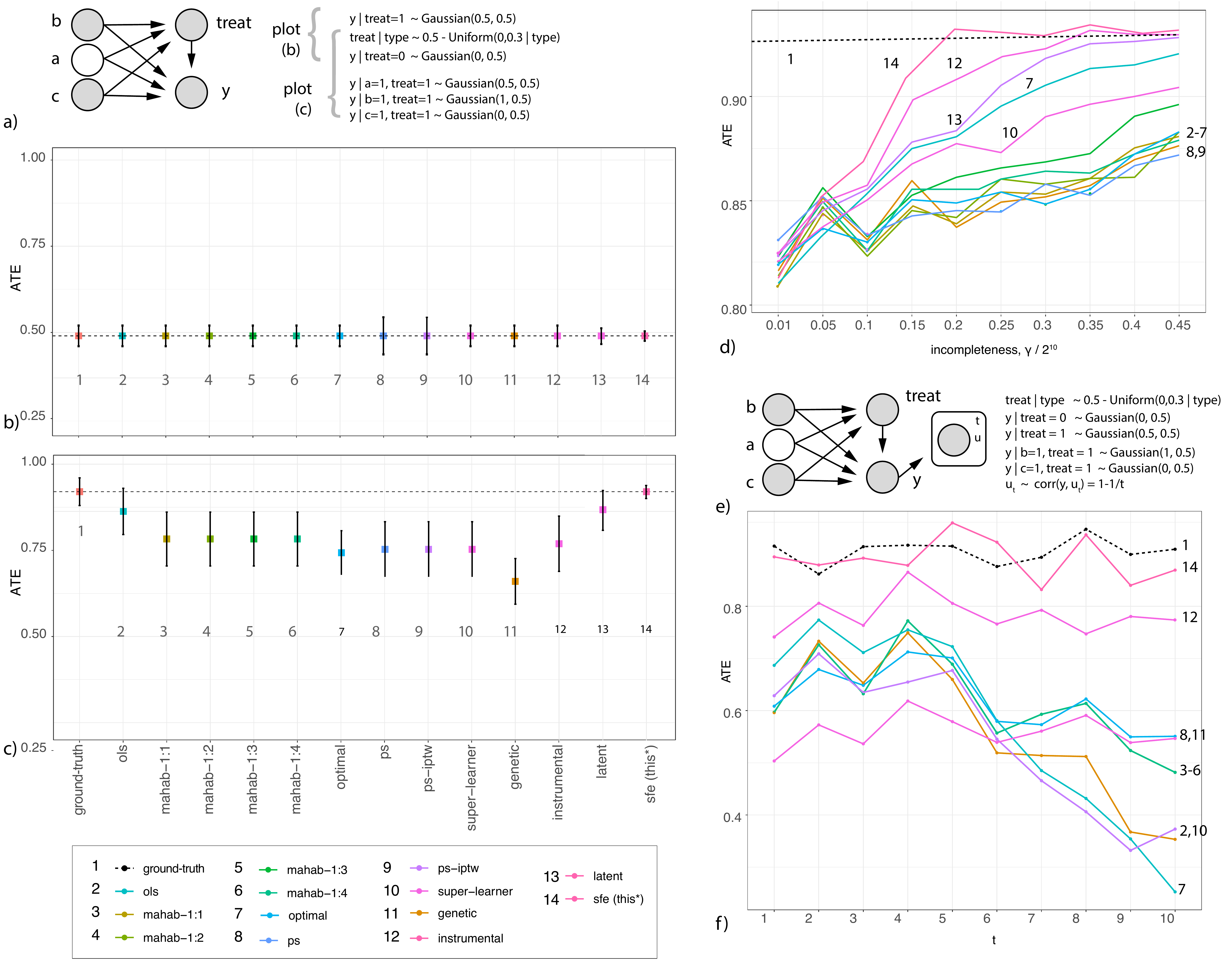}
\caption{\label{fig-simulation} Montecarlo simulations. \textbf{(a)} A priori Data-Generating Process (DGP) in graphical-model notation (gray variables are observed), population of 1000 individuals with 3 disjoint subpopulations $type=\{a,b,c\}$ with sizes $600$, $300$, and $100$, $treat$ is a treatment indicator and $y$ an outcome; \textbf{(b)} Average treatment effect (ATE) and Mean Squared Error (MSE, error bars) estimates from 1,000 simulations under observational conditions using several methods (columns), ground-truth is dotted, effects drawn from a Gaussian with mean and standard deviation $0.5$ for all subpopulations, treatment propensities for each subpopulation drawn from a uniform distribution over $[0.2,0.5]$, methods are: propensity-score matching (\textit{ps}), propensity-score with inverse probability of treatment weighting (\textit{ps-iptw}), mahalanobis covariate matching (\textit{mahab}) with 1-4 neighbor matchings, optimal matching (\textit{optimal}), ordinary least-squares (\textit{ols}), genetic balance optimization (\textit{genetic}) ~\citep{genetic}, SuperLearner ensembles (\textit{superlearner}) ~\citep{RefWorks:doc:5aa6f0c4e4b059d4ac147e0e,RefWorks:doc:5aa5b99fe4b0db66bcfcdb5c}, high-dimensional instrument selection (\textit{instrument}) ~\citep{ChernozhukovVictor2015PaPI,BelloniAlexandre2014IoTE}, latent causal variables deep-learning (\textit{latent}) ~\citep{NIPS2017_7223} and stochastic factorial estimation (\textit{sfe}), using eq.(\ref{eq-ate}); \textbf{(c)} simulations under heterogeneous conditions, effects drawn from Gaussians with mean $1.0$ for subpopulation $b$ and $0$ for $c$ (both with standard deviation $0.5$); \textbf{(d)} avg. factor ATE for simulations with 10 subpopulations (unitary effects) and increasingly complete samples $\gamma/2^{10}$ in respect to pairwise differences ($\gamma$ is the number of random sampled $\mathcal{C}^{10}$ edges, horizontal axis), \textbf{(e)} DGP with added $u_t$ variables (plate) that are increasingly correlated with $y$; \textbf{(f)} ATE estimates for simulations under confounding, heterogeneous and endogenous conditions, $1 \leq t \leq 10$. }
\end{figure}

\section{Results}
\subsection{Simulation}

We first test SFE in synthetic datasets. The design is similar to previous studies ~\citep{RefWorks:doc:5a514530e4b0e3e5a635f08a,Athey7353}. In the first simulations, the treatment effect is homogeneous. Each of 3 subpopulations are, however, randomly underrepresented in the treatment. Fig.\ref{fig-simulation}(a) shows the DGP in graphical-model notation and Fig.\ref{fig-simulation}(b) Average Treatment Effects (ATE) and their Mean Squared Errors (MSE) from several popular methods, which includes all methods in ~\citep{RefWorks:doc:5aa5b99fe4b0db66bcfcdb5c} plus 5 methods making use of Machine Learning (\emph{super-learner,genetic,latent,instrument,sfe}), Distance Metric Learning (\emph{genetic,latent,sfe}) and high-dimensional Econometric (\emph{instrument}) techniques\footnote{\label{note-methods} see ~\citep{RefWorks:doc:5aa5b99fe4b0db66bcfcdb5c} for further algorithmic details.}. The counterfactual outcomes $y^{\textrm{+treat}}$ and $y^{\textrm{-treat}}$ for each individual and $\textit{ITE}_i(treat)=y^{\textrm{+treat}}{-} y^{\textrm{-treat}}$ are known. In this setting, all methods recover the ground-truth (dashed line) with little bias. SFE has, however, the smallest MSE, even below the ground-truth's MSE, illustrating the advantage of using many pairs.

 \begin{figure}
\centering
\includegraphics[width=0.9\linewidth]{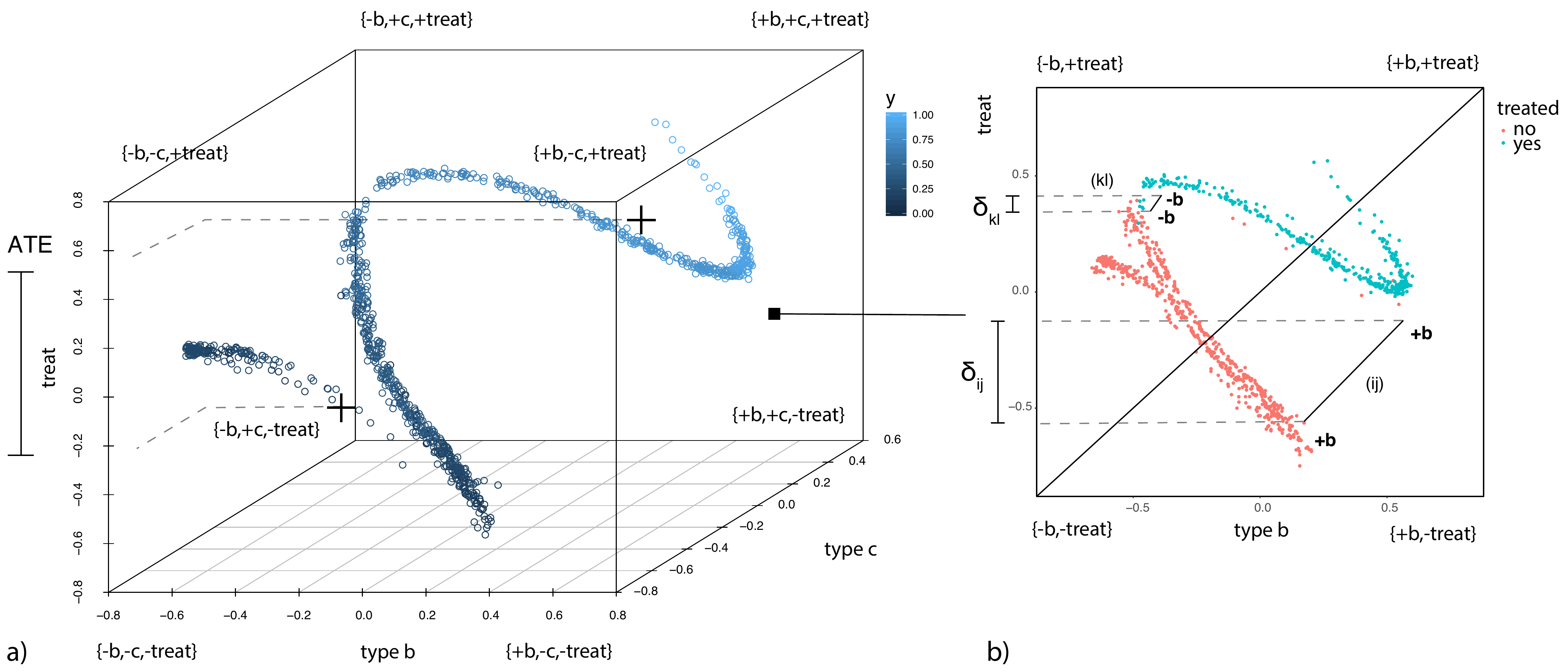}
\caption{\label{fig-simulation2} \textbf{(a)} Factorial subspace $\{b,c,treat\}$ of $T_y(X)$ for the simulations in Fig.\ref{fig-simulation}(b), pluses ($+$) show the (average) positions of treated and nontreated subpopulations, the $ATE$ corresponds to coordinate differences between these subpopulations on the $treat$ axis;  \textbf{(b)} subspace $\{b,treat\}$, treated individuals in green, nontreated in red, diagonal axis indicates the variables' combined effect, the ATE is an aggregation of individual effects, distances (solid lines) and their projections, $\delta_{ij}$ and $\delta_{kl}$, illustrate individual treatment effects for two example pairs: pair $(ij)$ with both individuals in subpopulation $b$, with $+b$ and distinct $treat$, and $(kl)$ with individuals not in $b$, with $-b$ and distinct $treat$.  }\label{fig:side}
\end{figure}

In the next simulations, we assume that subpopulations respond differently to the treatment. Subpopulation $b$ observes double the expected effect, whereas $c$ is immune. The resulting heterogeneity introduces significant biases in most estimates, Fig.\ref{fig-simulation}(c). Whereas most estimators become increasingly inaccurate as populations become more diverse, SFE continues to provide accurate and unbiased ATE estimates.

Fig.\ref{fig-simulation}(d) shows simulations with 10 subpopulations and samples that are decreasingly incomplete (i.e., approaching a Factorial Design). At each instant, a number $\gamma$ of the $2^{10}$ hypercube edges $\mathcal{C}^{10}$ is sampled uniformly without replacement. Treatment propensities are as before and effects are unitary. The figure shows mean $ATE$ estimates across factors (vertical axis) with increasing $\gamma/2^{10}$ (horizontal axis). The figure demonstrates that while the central assumption in the present work is factorial incompleteness, the assumption, in fact, impact other methods more severely. \textbf{Making such assumptions explicit is, we believe, one of the proposed representation's strengths.}  


Fig.\ref{fig-simulation2}(a) illustrates SFE's factorial representation. It shows a 3D subspace of the estimated space $T_y(X)$. Dots show individuals' positions and their outcomes (colors). Spatial differences in $T_y(X)$ reflect differences in outcomes $y$. The $ATE$ corresponds to differences in the $treat$ coordinate between treated and nontreated subpopulations, eq.(\ref{eq-ate}). Results demonstrate that SFE achieves lower MSE in homogeneous synthetic samples and can recover unbiased individual effects under heterogeneity. 
  
\begin{figure}
\centering
\includegraphics[width=1\linewidth]{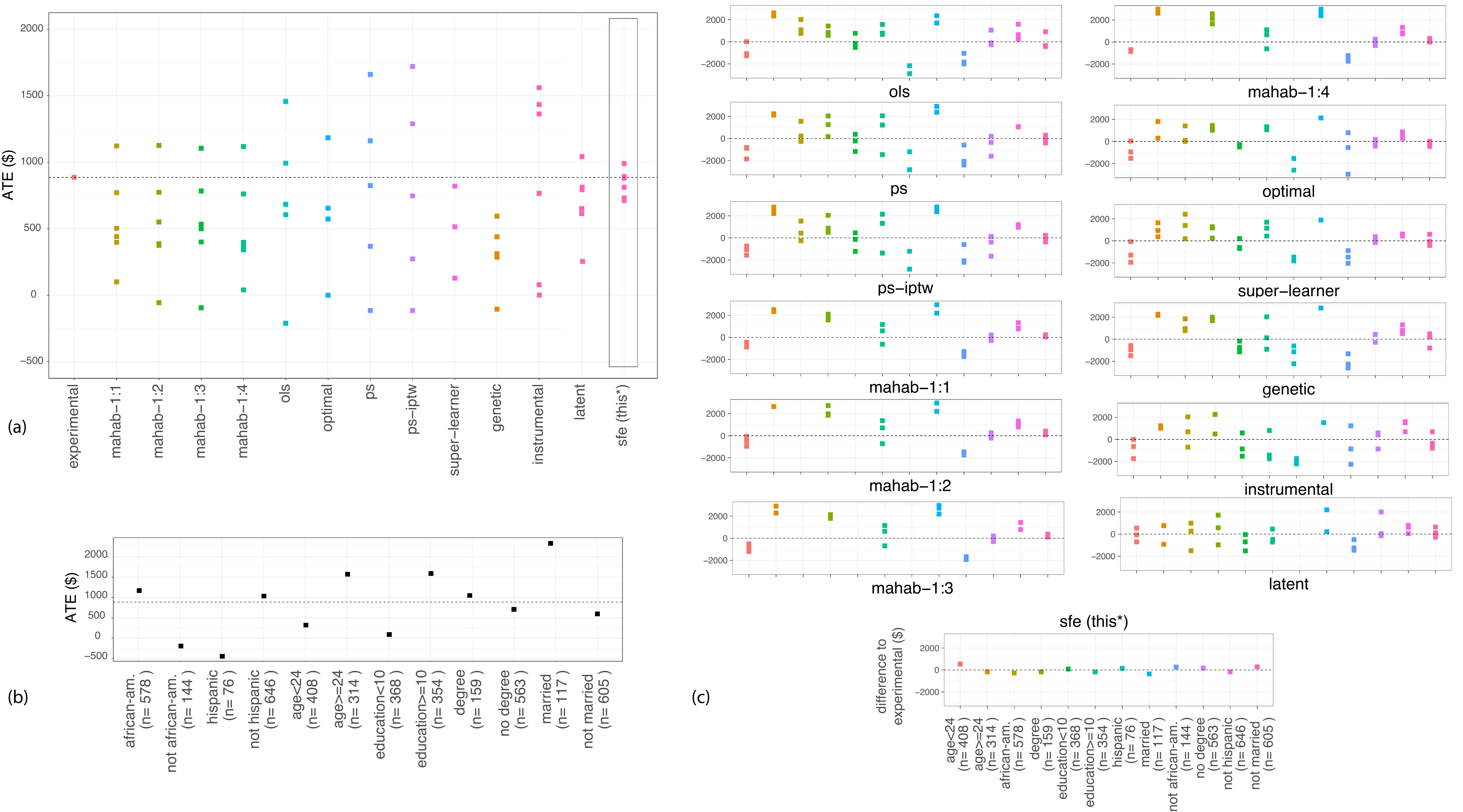}
\caption{\label{fig-lalonde} \textbf{(a)} Lalonde ~\citep{RefWorks:doc:5a5144d2e4b08e15c00ca6cf} reproduction, estimated $ATE$ for the National Supported Work (NSW) program with different methods (columns) over the 6 control surrogates drawn by Lalonde (dots), the dotted line shows the experimental estimate, variables as in ~\citep{RefWorks:doc:5a5144d2e4b08e15c00ca6cf}: $y$ is the worker's real post-program annual earnings (in 1981 dollars), $\mathcal{X}$ variables are workers' age, years of schooling, wage before entering the program, treatment status, race indicators (African-American, Hispanic) and whether the worker holds a high-school degree; \textbf{(b)} experimental effects of subpopulations (columns) in blocked samples (and their resulting sample sizes $n$); \textbf{(c)} estimated effects (difference from experimental, in dollars) for subpopulations (columns) according to each method, note that existing methods require 12 separate estimations, SFE represents subpopulations explicitly, making subpopulation-specific effects available from $T_y(X)$, eq.(\ref{eq-ate}), without re-estimation;  SFE estimates are closer to the ground-truth in both the Lalonde sample population and blocked subsamples.}\label{fig:side}
\end{figure}  
 
\subsection{National Supported Work (NSW) Program}
We now consider a real-world application: the NSW employment program (details in the Appendix B), where eligible applicants were randomized into treatment and control groups. In his seminal article ~\citep{RefWorks:doc:5a5144d2e4b08e15c00ca6cf}, Lalonde selected a subsample of the NSW participants and replaced its nontreated subgroup with samples from national surveys, leading to 6 distinct datasets. By doing so, he 'unbalanced' the NSW data (i.e., subpopulations' treatment propensities) - previously balanced by the NSW's experimental design. Lalonde then showed that observational methods failed to recover the experimental effect, a finding corroborated by later authors ~\citep{RefWorks:doc:5a514994e4b08e15c00ca73d, RefWorks:doc:5a514530e4b0e3e5a635f08a}. Subsequent research ~\citep{RefWorks:doc:5a5145dde4b0eeb35a49d118,RefWorks:doc:5a5145b6e4b01d3dd55629f7,RefWorks:doc:5a514994e4b08e15c00ca73d} showed, however, that in a more restricted sample (henceforth the 'DW' subsample) covariate matching and other methods recover the experimental effect. This small sample continues to be used to this day ~\citep{RefWorks:doc:5aa5b99fe4b0db66bcfcdb5c}. 

 Fig.\ref{fig-lalonde}(a) shows ATE estimates calculated by different methods (columns), using Lalonde's variables and sample restrictions, as well as the experimental (dashed line). Each dot is an estimate in one Lalonde sample. As in Lalonde's study, methods struggle to recover the NSW effect. In contrast, SFE estimates are consistently close to the experimental effect - within U\$$37$, well inside the experimental 95\% confidence interval. 
 
ATEs neglect that the NSW effect may differ across subpopulations.  Fig.\ref{fig-lalonde}(b) shows experimental effects for several subpopulations. Far from homogeneous, the program's effect was particularly large for older, married and relatively educated workers. Fig.\ref{fig-lalonde}(c) shows that, unlike other methods, SFE estimates these heterogeneous effects with negligible bias in all subsamples.

\subsubsection{Unknown or Unconsidered causes}

\begin{figure}
\centering
\includegraphics[width=0.8\linewidth]{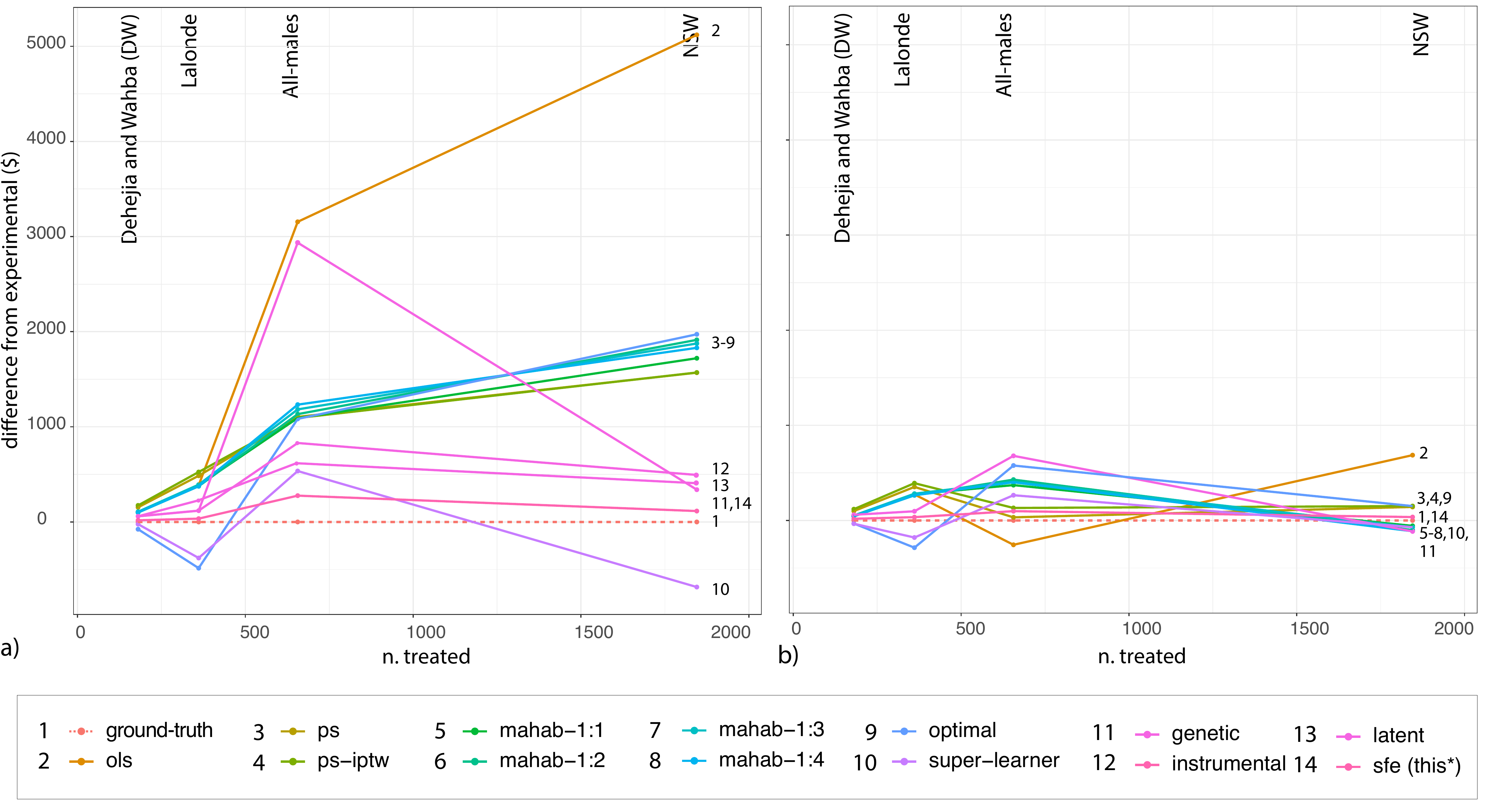}
\caption{\label{fig-scale} \textbf{(a)} ATE difference from experimental groundtruth (in dollars) using Lalonde's expert DGP and progressively fewer ex ante assumptions from previous studies ~\citep{RefWorks:doc:5a5144d2e4b08e15c00ca6cf,RefWorks:doc:5a5145b6e4b01d3dd55629f7,RefWorks:doc:5a514994e4b08e15c00ca73d,RefWorks:doc:5aa5b99fe4b0db66bcfcdb5c}, leading to the full NSW dataset (rightmost) with 1231 variables, 1923 treated and 8001 non-treated individuals; estimates become progressively harder for most approaches as sample restrictions are loosened with the exception of SFE; \textbf{(b)} ATE estimates with an automatically devised DGP (Appendix B details the procedure), yielding a DGP with variables: worker's work-ethic, race (African-American), previous school attendance, previous work, alcoholic drink consumption and location (New York city); all methods perform well throughout all samples using the DGP devised with SFE. }\label{fig:side}
\end{figure}  

 Both Lalonde and Dehejia and Whaba restricted the sample population and model variables. Fig.\ref{fig-scale}(a) shows results in 4 nested subsamples: DW, Lalonde, all males in the NSW and, finally, the full NSW dataset. This figure uses variables Lalonde picked based on his expert judgment. All methods perform well in the DW sample. However, their performance degrades as sample restrictions are relaxed. Fig.\ref{fig-scale}(a) and Fig.\ref{fig-lalonde}(c) suggest that SFE, in contrast, can estimate effects in heterogeneous populations, relieving the need for population selection. Can SFE also help determine which variables to include as causes? Fig.\ref{fig-scale}(b) show estimates with variables selected by SFE from the 1232 NSW variables (selection procedure details in the Appendix B). Using this alternative set of 5 variables improves SFE performance. More surprisingly, it also improves other estimators. Using these variables, all observational methods approach the ground-truth (matching methods, in particular), even as sample restrictions are removed. This suggests that the autonomously identified DGP approximates the true DGP more closely than the one derived from expert judgment. 

\begin{figure}
\centering
\includegraphics[width=0.8\linewidth]{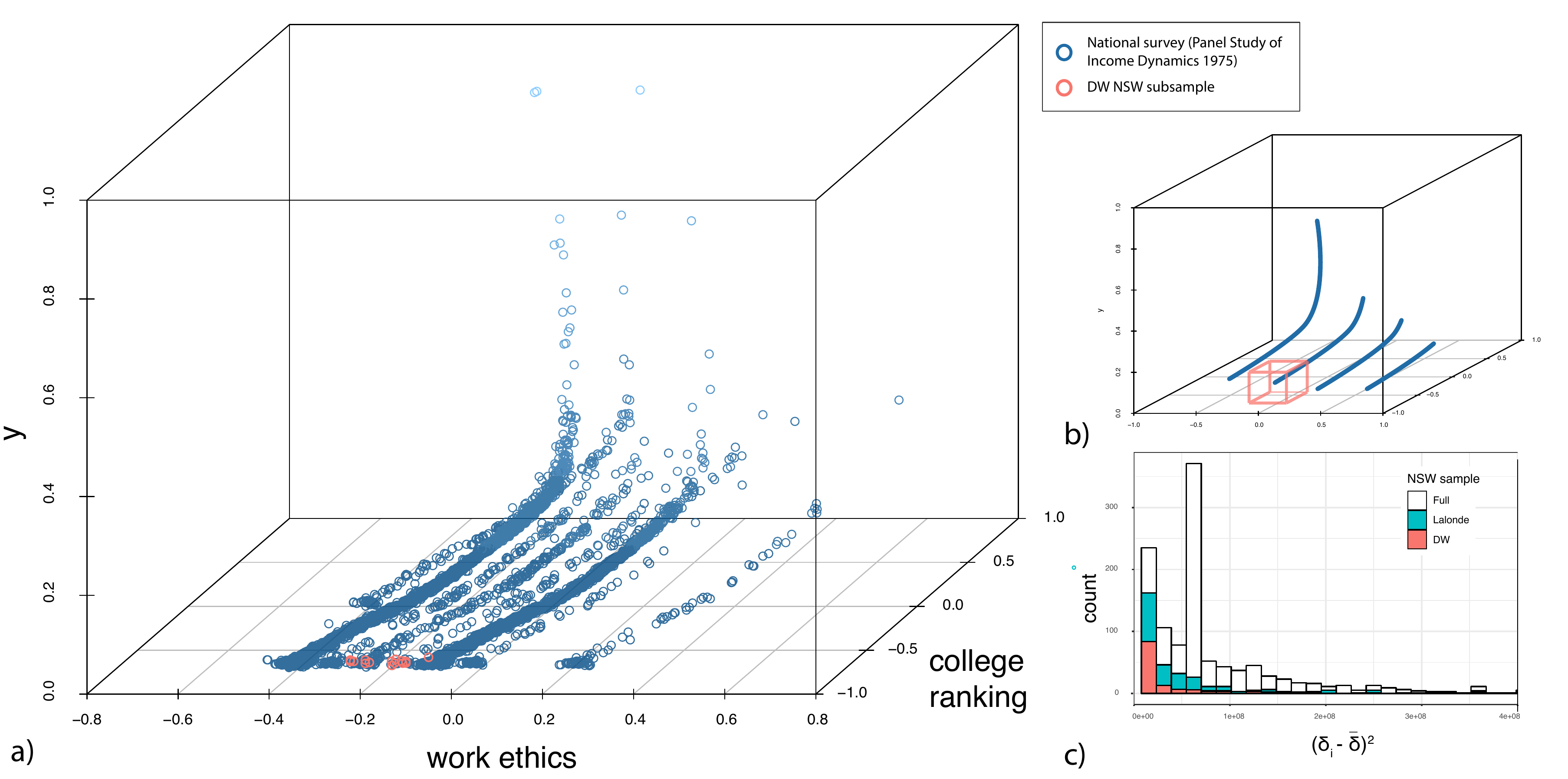}
\caption{\label{fig-scale-viz} \textbf{(a)} Subspace of $T_y(X)$ consisting of two variables (ranking of attended college and work-ethics) versus outcome $y$ (wages), variables selected after running SFE on a matched NSW-National survey (Appendix B describes the procedure, dataset has 441 variables, 8001 individuals), survey respondents are blue and individuals in the Dehejia and Wahba (DW) sample are red; \textbf{(b)} illustration of diminishing exponential returns across subpopulations and the area of concentration of the DW sample (small cube); \textbf{(c)} $({\delta}_i - \bar{\delta})^2$ histogram for treated individuals in DW, Lalonde and the full NSW experimental samples, $\delta_i$ are before-after individual outcome differences, $0 < i \leq n$, and $\bar{\delta}$ their mean in each sample; results illustrate how the DW sample contains a highly homogeneous subpopulation.}\label{fig:side}
\end{figure}  

But why did all observational methods perform well in the DW sample? To explain this, Fig.\ref{fig-scale-viz}(a) plots individuals' coordinates on the two variables with the highest ATEs in a NSW-National Survey matched dataset: work-ethics and college ranking.  It illustrates how wages increase exponentially with college ranking, while, at the same time, this relation varies with individuals' work-ethic. The figure shows in red the matched DW subsample. This suggests that the reason why observational methods recover effects in the DW sample with apparent ease is that this subsample consist of individuals with high effect homogeneity.

\subsection{Discussion}
Which DGP was selected by SFE? Table \ref{table-vars} in the Appendix B lists the 20 variables with the largest effects against 20 selected by a traditional model-selection algorithm, a regularized (LASSO) regression. The two lists are very different. Among the variables selected by a LASSO are \emph{alimony money received}, \emph{unemployment in past two years}, \emph{money received from training}, \emph{money from social security}. Several of these reflect consequences, or just components, of an individual's income. In contrast, the top variables selected by SFE are \emph{work-ethics}, \emph{race (African-American)}, \emph{NSW treatment}, \emph{recent school attendance} and \emph{recent employment}. All these variables are arguably connected to causes of income differences such as education, work attitudes and discrimination. 

These results suggest that SFE may also shed light on the direction of causation. To explore this further, we revisit the earlier simulations, adding variables that are consequences, not causes, of the outcome variable. We progressively add 10 variables $u_t$, $0< t\leq 10$,  which are increasingly correlated with $y$, with expected Pearson correlations $1-\frac{1}{t}$. Fig.\ref{fig-simulation}(e) shows the expanded graphical-model and Fig.\ref{fig-simulation}(f) ATE estimates. For most methods, even a small numbers of consequences significantly biases estimates. In contrast, SFE estimates remain unbiased. 

We introduced SFE in this article, a computational tool for nonexperimental causal effect estimation which we compared to several estimators from the Statistics and Machine Learning literature. SFE allows researchers to represent nonexperimental data as incomplete factorial designs, eq.(\ref{eq-goal}). We have shown that, as result, it can recover the ground-truth in synthetic data and in Lalonde's seminal setting - estimating causal effects with less bias and error than alternatives. We have also shown  effect estimates at the individual level and in the entire nationwide NSW program, not relying on ex ante model and population selection criteria, outperforming estimates that used expert specifications. A more abstract goal was to demonstrate that the troves of data on pairwise treatments and confounders in common nonexperimental data can be very useful when estimating causal effects.  Many fields, from Medicine to the Social Sciences, face new realities where historical data is increasingly accessible and new data is constantly accumulating. The tool could enable new uses for such data in scientific investigation.



\bibliography{arxiv}

\newpage

\appendix
\setcounter{section}{0}
\section{Objective Function}

In this section, we devise eq.(\ref{eq-sol-main}) for binary observed variables. We consider an extension for continuous variables in Sect.\ref{sect-sm}. 

\subsection{Treatment Likelihood}\label{sect-fact-dist}
Let's first define requirements for a pair of individuals $(ij)$ to represent an univariate treatment $\{a\}$ with certainty, $a \in \mathcal{X}^m$. A first requirement relates to the treated variable $a$ itself\footnote{for short, we use $x$ to refer to both Boolean vectors and set variables (i.e., the set of variables with value +1).}. The requirement is that $x_i(a) \cdot \neg x_j(a)=\mathbbm{1}$, where $\neg$ and $\cdot$ are the boolean NOT and AND operators and $\mathbbm{1}$ is a $m$-sized vector with all $+1$ values. A second requirement relates to other variables, $b \neq a$. The requirement is that these variables are either also treated, $x_i(b) \cdot \neg x_j(b)=\mathbbm{1}$, or common, $x_i(b) \cdot x_j(b)=\mathbbm{1}$,  between $i$ and $j$.  

With these requirements, we will define individuals' positions, $x_i$, as random observations of factorial runs. The norm of vectors, $\vert x_i \vert$ and $\vert x_j \vert$, relate to the likelihood of treatment and angles, $\theta_{ij}$, to observed confounding conditions among pairs of individuals. Their dot-product, $\inner{x_i}{x_j} = \vert x_i \vert \vert x_j \vert \cos \theta_{ij}$, will reflect both factors and become a key element in the optimization. 

 \begin{figure}   
     \centering
{\includegraphics[width=0.3\textwidth]{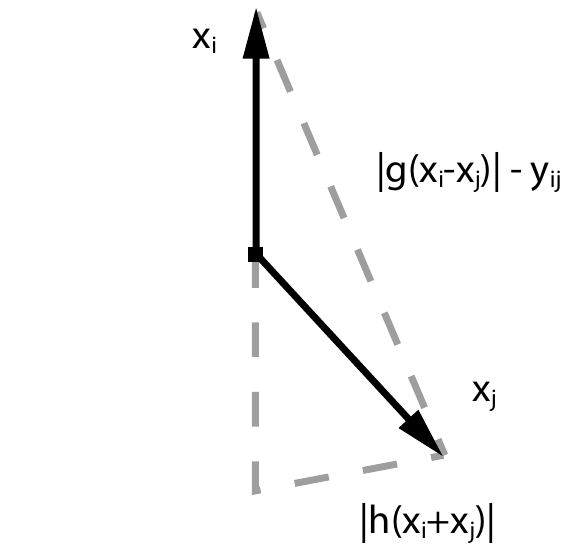}}
{\caption{\label{fig-triangle} example vector pair and the relationship among its dot-product, vector sum and vector difference, as well as, functionals - $g(x)$ and $h(x)$ - and outcome pairwise differences, $y_{ij}$.}  }
\end{figure}

 Before formulating this relationship in detail, reconsider eq.(\ref{eq-main}). Let $g(x) = \max(0,x)$ which makes nonpositive coordinates in $x$ zero\footnote{this function is often called a rectifier and is currently the most popular activation function in deep neural networks.\label{foot-rectifier}}. We can decompose an individual pair into vectors for their difference, $x_i{-}x_j$, and sum, $x_i{+}x_j$. Due to the sign convention, the first contains treated coordinates and the second non-treated coordinates. The dot-product relates the sum of the two vectors geometrically when $g(x){=}h(x)$. According to eq.(\ref{eq-main}), this corresponds to the assumption that the variance is proportional to the expected effect of non-treated variables, i.e., the expected amount of confounding. The relationship leads to a general least-squares solution (considered in further detail below), where we minimize a residual, $y_{ij} - \vert g({x}_i{-}{x}_j) \vert$, and a penalty, $\vert h({x}_j{+}{x}_j) \vert$. Notice that $\vert g({x}_i{-}{x}_j) \vert$ is also a distance. Fig.\ref{fig-triangle} sketches the (distance) residual and cost for an example pair.  Letting $x^{\prime} =g(x)$, the Law of Cosines leads to   

\begin{equation}
\begin{split}  
& \Big(y_{ij} -\vert {x}^{\prime}_i - {x}^{\prime}_j \vert \Big)  + \vert {x}^{\prime}_i + {x}^{\prime}_j \vert  ,\\
&=  \Big( y_{ij} - \vert {x}^{\prime}_i \vert^2  - \vert {x}^{\prime}_i \vert^2 + 2 \inner{{x}^{\prime}_i}{{x}^{\prime}_j} \Big) \\
& \qquad+ \vert {x}^{\prime}_i \vert^2  + \vert {x}^{\prime}_i \vert^2 + 2 \inner{{x}^{\prime}_i}{{x}^{\prime}_j},\\
&=  y_{ij} + 4 \inner{{x}^{\prime}_i}{{x}^{\prime}_j} ,\\
\end{split}
\end{equation}

\begin{figure}
\centering
\includegraphics[width=0.6\linewidth]{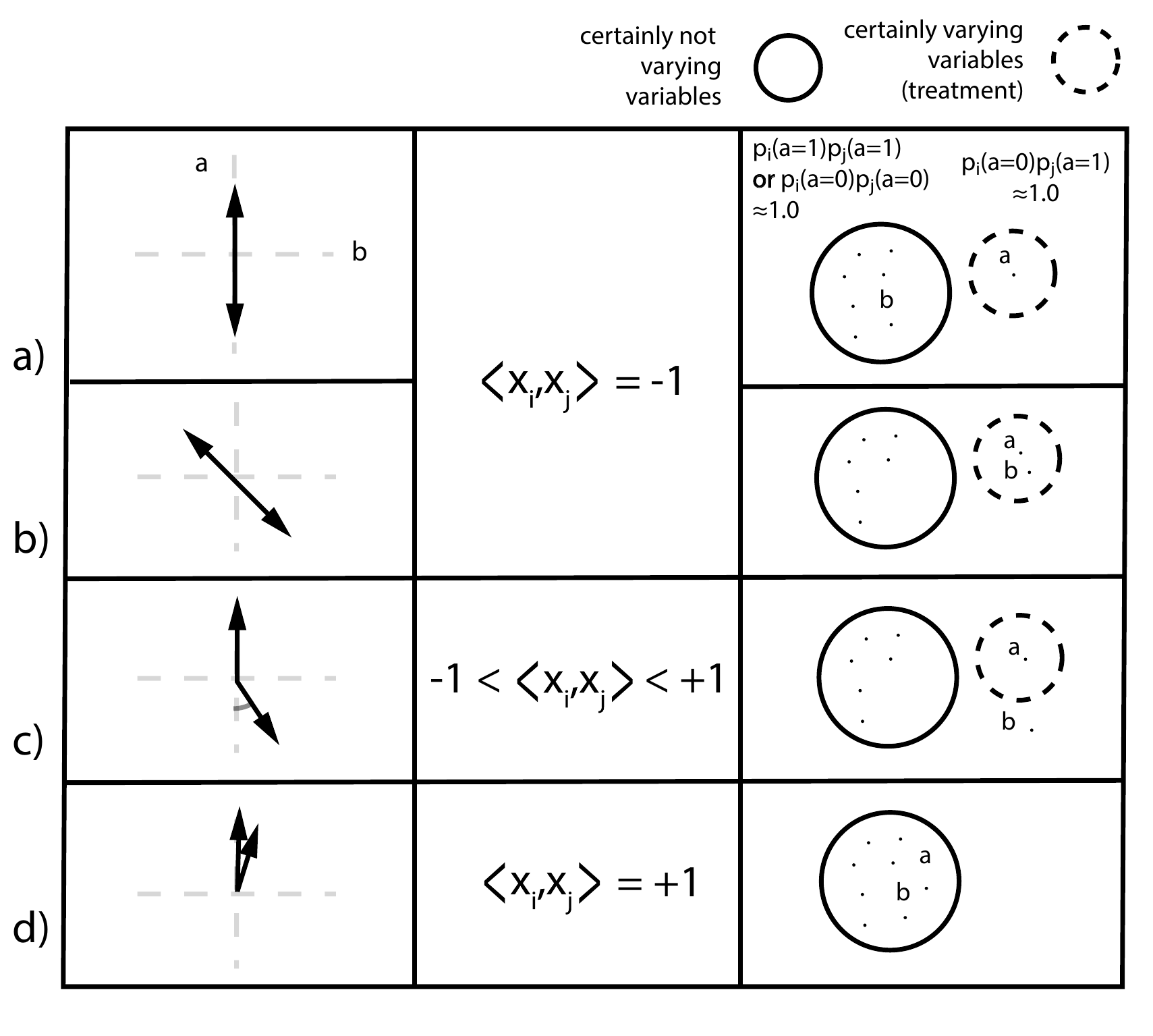}
\caption{\label{fig-cases} Representative pairings of individuals (leftmost column) and their related dot-product (middle column) and treatments as Venn diagrams (rightmost column): \textbf{(a)} univariate individual-level treatment with no observed confounding risk; \textbf{(b)} multivariate treatment with no confounding risk; \textbf{(c)} univariate treatment with confounding risk, \textbf{(d)} no treatment.}
\label{fig:frog}
\end{figure}

Let's then define the probability $p_y({x}_i,{x}_j)$ and its relation to the dot-product in further detail. We defined a variable $a\in \mathcal{X}^m$ as being under a factorial treatment when a variable is treated and all other variables are either common or also treated. Fig.\ref{fig-cases}(a-d) (3rd column) depicts these conditions as Venn diagrams for the cases in Fig.1 (main article). The dot-product in the $2^m$-dimensional Boolean vector space ~\citep{boolfunc} has the interpretation

\begin{equation}\label{eq-expected-dot0}
  \inner{z_i}{z_j} = \frac{1}{2^m} \sum_{a=1}^m z_i(a)z_j(a) =  E_U[z_i\cdot z_j]  
\end{equation}

where $z \in \{\texttt{-}1,\texttt{+}1\}^m$ and the expectation is taken uniformly over all $a \in \mathcal{X}$. The dot-product indicates the expected number of common variables between vectors. It also defines the $\ell_2$-norm $\vert z \vert = \sqrt{\inner{z}{z} } = \sqrt{E_U[z^2] }$. We consider, instead, 

\begin{equation}\label{eq-expected-dot}
  \inner{z_i}{\neg z_j} =  E_U[z_i \cdot \neg z_j],  
\end{equation}

which indicates the expected number of treated variables between vectors, when $\mathcal{X}$ variables are uniformly distributed (notated $U$). Non-uniform distributions and continuous treatments are considered in Sect.\ref{sect-sm}. Due to the $\{\texttt{-}1,\texttt{+}1\}$ sign convention, the product in the standardized covariate space $X$ leads to the relation    

\begin{equation}\label{eq-dot-likelihood}
 E_{U} [x_i \cdot \neg x_j] =  \begin{cases}
     -\inner{ {x}_i}{{x}_j},& \text{if } \langle {x}_i,{x}_j \rangle \leq 0\\
    0,              & \text{otherwise}
\end{cases}. 
\end{equation}

When $\inner{{x}_i}{{x}_j}=-1$, the probability of drawing a treated variable when comparing $i$ and $j$ is 1.0 - i.e., $x_i \cdot \neg x_j {=}+1, \forall a \in \mathcal{X}$. We also associate the  pair with a treatment size, $\phi^{cx}_{ij}$, and a possible confounding risk in relation to the remaining $n{-}2$ individuals, $\phi^{bl}_{ij}$. We consider these in a Bayesian framework next.

\subsection{Sample Balance and optimization}
We now turn to conditions $\phi_{ij}$. We will represent these conditions geometrically, while, at the same time, relating them to the density $p_y({x}_i,{x}_j)$. Eq.(\ref{eq-main}) implies the following likelihood over expected effects\footnote{The typical likelihood notation $\mathcal{N}[ y | \mu, \sigma^2]$, for an observation $y$ with mean $\mu$ and variance $\sigma^2$, is used.}:

\begin{equation}\label{eq-likelihood}
\begin{split}  
\prod_{i,j} \mathcal{N}[ y_{ij} \; |   \; p_y({x}_i,{x}_j) f({x}_i{\ominus}{x}_j),  h({x}_j {\ominus} {x}_i) ].
\end{split}  
\end{equation}

To consider also learning from pairs with balanced treatments, we introduce a Gaussian prior $\mathcal{N}({x}_i{\ominus}{x}_j \; |  0, \phi^{bl}_{ij})$ for the probability $p_y({x}_i,{x}_j)$, where $\phi^{bl}_{ij}$ is a strictly positive scalar for each pair. If not a factorial treatment, the probability that the pair represents the treatment ${x}_i{\ominus}{x}_j$ depends on the likelihood that non-common factors in the pair (2-sample) are balanced in the remainder of the sample. 

Combining the likelihood in eq.(\ref{eq-likelihood}) with the Gaussian prior, we obtain

\begin{equation}\label{eq-likelihood2}
\begin{split}  
\prod_{i,j} \mathcal{N}[ y_{ij} \; |  \; p_y({x}_i,{x}_j) f({x}_i{\ominus}{x}_j),  \phi^{cx}_{ij} ]\times \mathcal{N}[ {x}_i{\ominus}{x}_j \; |  \; 0,  \phi^{bl}_{ij} ],
\end{split}  
\end{equation}

where $\phi^{cx}_{ij} = h({x}_i{\ominus}{x}_j)$ is the pair's variance. This formulates Bayesian priors for conditions $\phi_{ij}$ from eq.(\ref{eq-main}) in a way similar to a Tikhonov regularization ~\citep{regular}.

Considering a single position ${x}_i$ and treatment $v$, our goal is to transform ${x}_i$ such that $\vert \mathbf{x}_i - \mathbf{x}_j \vert^2 = p_y({x}_i,{x}_j)f({x}_i{\ominus}{x}_j)$ for $0 < j \leq n$. Combining likelihoods in eq.(\ref{eq-likelihood2}) and eq.(\ref{eq-dot-likelihood}), taking logarithms and dropping constants we arrive at the objective

 \begin{equation}\label{eq-sol}
\begin{split}
\Gamma({x}_i) &=  \min_{{x}_i} \sum_{j=1}^n (1+\hat{\phi}_{ij}^{cx})[ \inner{{x}_i}{{x}_j} + y_{ij}]^2 + \hat{\phi}_{ij}^{bl} \inner{{x}_i}{{x}_j}^2 +b_i,\\
 &\hat{\phi}_{ij}^{cx} = \frac{\vert {x}_i + {x}_j \vert}{m}, \\ 
 &\hat{\phi}_{ij}^{bl} = \vert \frac{1}{n}\sum_{k=0}^n \inner{{x}_k}{{x}_i+{x}_j} \vert,\\
 & y_{ij} = \max(0, y_i - y_j).   
\end{split}
 \end{equation}

We consider the overall objective function first, then pairwise penalties estimates, notated $\hat{\phi}_{ij}$, followed by the intercept $b_i$ and outcome differences $y_{ij}$. If we minimize eq.(\ref{eq-sol}) with respect to the $m$-sized vector ${x}_i$ we get a maximum a-posteriori likelihood estimate for individuals' positions. The objective function argument is an individual's position ${x}_i$ (rowspace vectors) and not factor positions (column space vectors). More specifically, eq.(\ref{eq-sol}) leads to an iterative gradient minimization procedure for each individual, $\mathbf{x}^{t+1}_i = \mathbf{x}^t_i -\eta\Gamma(\mathbf{x}^t_i)$, with $\eta$ as learning rate and $\mathbf{x}_i^{t=0} = {x}_i$. Considering the entire sample population, we iteratively minimize their gradient sum, $\sum_{i}[\mathbf{x}^t_i - \Gamma(\mathbf{x}^t_i)]$. Both Statistics and Machine Learning researchers have considered the problem of minimizing an objective function in the form of a sum of gradients.  We use a Stochastic Gradient Descent (SGD) ~\citep{RefWorks:doc:5a518de8e4b08e15c00cac6a} which samples a subset of summand functions at every step and has found wide-spread use in Machine Learning ~\citep{RefWorks:doc:5a51905fe4b01d3dd556309f,RefWorks:doc:5a518fd7e4b0eeb35a49d719}. The scheme allows us to consider billions of observation pairs when estimating effects. We discuss other implementation details in Sect.\ref{sect-implementation}. The resulting optimization transforms the original space $X$ into $T_y(X)$. It transforms treatment vector differences, until they reflect difference in outcomes that approximate, according to the defined costs, those that would be observed in factorial experiments. 


Eq.(\ref{eq-sol}) defines $y_{ij}$ as nonnegative outcome differences. The scalar term $b_i$  is an individual's intercept with expected zero mean that is also minimized. Terms $\hat{\phi}^{bl}_{ij}$ and $\hat{\phi}^{cx}_{ij}$ reflect difference-of-means balance and treatment size conditions for pairs of individuals $i$ and $j$. Pairs with both zero penalties (balanced and univariate treatments) reproduce, according to the previous assumptions, factorial or randomized treatments. In this case, $\inner{{x}_i}{{x}_j}$ is made to reflect $y_{ij}$.  

Calculations will run over thousands of iterations for large observation matrices $X$. Therefore, it is important to define simple penalties $\phi_{ij}$. We defined treatments by dividing pairs' variables into treated and common variable subsets. With the $[\texttt{-}1,\texttt{+}1]$ sign convention, the vector ${x}_i + {x}_j$ has non-zero values for non-treated variables. In eq.(\ref{eq-sol}), the penalty $\hat{\phi}^{cx}_{ij}$ is therefore a normalized estimate for the number of non-treated variables. Non-treated (i.e., non-zero) coordinates in $x_i+x_j$ can confound outcome effect observations, $y_{ij}$, Fig.\ref{fig-triangle}(c). We did not deem pairs under these conditions as necessarily unsuitable for estimation. Instead, we considered that the pair has coordinates that need to be balanced in individuals $k$ that do not belong to the pair, $k\neq i,j$. For an out-of-pair individual $k$, $\inner{{x}_k}{{x}_i{+}{x}_j}$ is the projection of that individual's vector ${x}_k$ onto ${x}_i{+}{x}_j$. The penalty $\hat{\phi}^{bl}_{ij}$ is a sum of such projections from all other $n-2$ individuals (signed, due to the same convention). Orthogonal vectors have null projections, and, balanced vector-sets have null sums.  


\section{Supporting Material}\label{sect-sm}

\subsection{Continuous Treatments}\label{sect-continuous}
We can also use the previous method with continuous variables, when it is assumed that there is uncertainty over the intensity of treatments. This can be carried out either by extending $p_y(x_i,x_j)$ directly or by considering a third Bayesian factor for treatment intensity in eq.(\ref{eq-likelihood}) (together with treatment balance and size). We consider the former.  In Computational Learning Theory, a \emph{product distribution} ~\citep{product-distr1,product-distr2} is a distribution over $\{0,1\}^m$ which generalizes the relationship in eq.(\ref{eq-expected-dot0}) to the non-uniform case. We define a distribution 

\begin{equation}\label{eq-prod-distribution}
\begin{split}
 \mathcal{D}_{ij} &= \prod_{p_i(\texttt{+}a)\leq p_i(\texttt{-}a)}p_j(\texttt{+}a)\prod_{p_i(\texttt{+}a)> p_i(\texttt{-}a)}p_j(\texttt{-}a),\\
\end{split} 
\end{equation}

where $p_i(\texttt{+}a)$ is the probability that individual $i$ has factor $a$ and $p_i(\texttt{-}a)$ that he doesn't, $a \in \mathcal{X}$. The first therefore indicates certainty of positive treatment status and the second of negative. Any continuous value in between corresponds to individuals with uncertain treatment statuses. 

This generalization preserves the relationship in eq.(\ref{eq-dot-likelihood}), where $x_i$ becomes the observational random vector with ${x_i}(a) = p_i(\texttt{+}a) - p_i(\texttt{-}a)$. In this case, the dot-product reflects the expectation over $\mathcal{D}_{ij}$ instead of $U$ ~\citep{product-distr1}. With a single observation per individual, a simple way of obtaining these vectors is unity-base normalizing (i.e., feature scaling) the observation matrix $X$ and assuming any applicable prior for values in-between, $[\texttt{-}1,\texttt{+}1]$. This makes maximum and minimum correspond to treated and nontreated statuses, with intermediary treatments having, for example, exponentially decreasing intensities. 

\subsection{Implementation}\label{sect-implementation}

The method can be carried out for all individuals in parallel with matrix operations. For results in this article, we first unity-base normalize $X$, 

\begin{equation}\label{eq-t0}
  T^{0}(X) = 2(X - X_{\min}) \oslash (X_{\max}-X_{\min}) - 1.0,
\end{equation}

where $X_{\min}$ and $X_{\max}$ are $m \times n$ matrices with per-column maximum and minimum values of $X$ and $\oslash$ is the element-wise (schur) division.  

\begin{algorithm}
 \caption{Space $X_y$ estimation.}\label{method}
      \SetAlgoLined
     \KwData{$X,y$}
     \Parameter{$\eta$ (learning rate)}
     \KwResult{$X_y = T_y(X)$ }
     calculate matrices $X_{\min}$ and $X_{\max},$\\
     calculate matrix $Y_{ij}$  \tcp*[r]{$y_{ij}= max(0,y_i-y_j)$}
     $X_0:=T^{0}(X)$\\
     \While{approximate minimum not obtained}{
      randomly shuffle $m$ examples from $X$;\\
      \For{$i := 1$ \KwTo $m$}{
         calculate matrices $\Phi^{cx}_{ij},\Phi^{bl}_{ij}$\\
         $A := \frac{1}{m} X_i*X_i$\\
         $A := ( A \otimes \Phi^{cx}_{ij} -Y_{ij})^2 + A^2 \otimes \Phi^{bl}_{ij}$ \tcp*[r]{$\otimes$ is the schur product}
         $\nabla X := \frac{1}{m} (A*X^\mathrm{T}_{i-1})^\mathrm{T}$\\ 
         $X_{i}:=X^\mathrm{T}_{i-1}-\eta \nabla X$ \\
      }
      $X_0 := X_m$ 
     }
     \Return $X_{0}$

\end{algorithm}

Eq.(\ref{eq-t0}) calculates the initial space $T^0(X)$. Subsequent transformations are performed by gradient descent over the sample population. The resulting method is summarized in Algorithm.\ref{method}.  All results in this article use 10,000 iterations and a learning rate of $\eta=0.025$. An optimized C++ version estimates a space $T_y(X)$ for the NSW dataset in under 5 minutes on a Macbook laptop. 

\subsection{NSW Study Details}\label{sect-missing}

\begin{table*}[ht!]
\centering
\caption{\label{table-vars} 20 top-variables according to $T_y(X)$ and LASSO regression in the complete NSW, ordered by estimated effect or correlation. Bolded variables were selected in the employed model selection criteria (see Sect.\ref{sect-missing}).}
 \begin{tabular}{ |S |L | } 
 \hline
 select by & NSW Variables \\
 \hline
 Eq.(\ref{eq-ate}) & \textbf{importance getting ahead in life: hard work}, \textbf{african american}, \textbf{treatment}, \textbf{not in school last 6 months}, \textbf{worked \textless 40 hours last 4 weeks}, \textbf{alcohol and hard-liquor consumption}, target NSW group (AFDC, ex-offender, ex-addict, youth, other), importance getting ahead in life: education, age \textgreater 18, technical eligibility flags, \textbf{site location (New York City)}, looking for work last in the last 4 weeks, site location (Philadelphia), no job in the last 6 months, searched for jobs directly from employer, earing less than program minimum (eligibility criteria), other gross eligibility flag (not revealed in the public file), youth group, employed \textless 9 months last year, program assignment year 1976  \\  
 \hline
 LASSO regression, $\beta$ & amount of alimony money, unemployed in 8th pre-program month (timeline), amount of money from training, NSW program, amount of money from social security, amount of SSI dollars, consumed drug other than marijuana, amount of money from alimony/child support, money from welfare, receive workman's compensation,  amount money from other welfare programs, gender, gender (eligibility flag), any money from workman's compensation, other gross eligibility flag (not revealed in public file), ever gone to school, how related to the person living with the participant (one of max. 12 persons living with participant), number of children, holds bachelor's degree, amount of money from AFDC program, how old is relative (one of max. 12 persons living with participant)  \\ [1ex] 
 \hline
\end{tabular}
\end{table*}

The NSW was a 1970s subsidized work program, running in 15 cities across the US for 4 years. It targeted individuals with longstanding employment problems: ex-offenders, former drug addicts, recipients of welfare benefits, and school dropouts. At the time of enrollment, each NSW participant was given a retrospective baseline interview, generally covering the previous two years, followed by up to four follow-up interviews scheduled at nine-month intervals. Survey questions covered demographic and behavior topics such as age, sex, race, marital status, education, number of children, employment history, job search, job training, mobility, housing, household, welfare assistance, military discharge status, drug use and extralegal activities. Most questions were objective and probed for specific information loosely around the previous themes (e.g., 'what kind of school are you going to? 1 = high school, 2 = vocational, 3 = college, 99 = other', 'was heroin used in the last 30 days?' etc.) Some questions were subjective (e.g., 'tell me how important each one is to you. knowing the right people, education, luck, hard work, ...')

To assemble control surrogates for the NSW, Lalonde used the \emph{Panel Study of Income Dynamics} (PSID), a household survey, and the Westat's matched \emph{Current Population Survey-Social Security Administration} file (CPS). He drew 3 subsamples from each the PSID and CPS (6 in total). Control groups had 450, 550, 726, 2666, 2787 and 16289 individuals\footnote{thus approximately 100K-200M treatments.}. Lalonde ex ante assumptions for the NSW, PSID and CPS regarded mainly participants' assignment date, gender, retirement status, age and prior wages. DW added further assumptions regarding prior wages for the NSW and used Lalonde's control groups. 

For the 'missing causes' study, we first estimated a model $T_y(X)$ where 

\begin{equation}
\mathcal{X}=\{all\: 1231\: variables\: in\: the\: NSW\: dataset\}.
\end{equation}

We use the same outcome variable $y$ as Lalonde, post-program annual earnings (in 1982 dollars). While using all NSW variables (i.e., the answer to every survey question), we only restrict them in one way. The restriction doesn't reduce the participant and variable counts. We ignore any variable values that are negative or '99', taking them as omitted - these values are then mapped to $x(a) = 0$ values according to eq.(\ref{eq-t0}). These correspond to unknown, not responded or exceptional values in the survey. We assume SFE should be able to handle other types of entry. Most variables are binary and naturally normalized to $[\texttt{-}1,\texttt{+}1]$ according to eq.(\ref{eq-t0}). Other variables are coded to reflect a spectrum (e.g., 'even though the 1000 could result in arrest, how likely is that you would take the chance? 1 = very likely, 2 = somewhat likely, 3 = somewhat unlikely, 4 = not likely at all') and they are accordingly mapped to $[\texttt{-}1,\texttt{+}1]$. Continuous and count variables are similarly linearly normalized to fit the interval (with maxima mapped to +1 and minima to -1). Following Lalonde's protocol, we 'annualized' the data. Participants' assignment date and location are not in the NSW data (only the participant's relative time in the program). Lalonde recovered site locations and assignment years by matching reported sites' unemployment to unemployment in \textit{Earnings and Employment} magazines. This is described in detail in ~\citep{RefWorks:doc:5a5144d2e4b08e15c00ca6cf}. Annualization allowed Lalonde to select only the 1975 participants. We, instead, added participants' estimated year of assignment and program site location as extra variables. 

Table \ref{table-vars}(first row) lists the 20 variables with largest $ATE(a)$ in the NSW, ordered by effect size. We also show the output of a LASSO estimator, as a more typical model selection procedure, Table \ref{table-vars}(second row). The NSW treatment indicator appears as the third most influential variable, but it doesn't appear in the list of variables selected by LASSO, Table \ref{table-vars}. Effect sizes relate to norms in $T_y(X)$ and dependence among variables to angles\footnote{remember that the cosine of angles between pairs of vectors in standardized datasets correspond to their Pearson correlation.}. To compare a SFE-devised DGP with Lalonde's, we next select a variable set of the same size as the one used by Lalonde. We choose effective and non-redundant causes. Let then $M$ be a set of unrelated variables $M \subset \mathcal{X}$, where $\mathcal{X}$ is the previous set of 20 effective variables. Furthermore, let $M^{k=K}(\mathcal{X})$ be a subset of $\mathcal{X}$ with $K$ variables and 

\begin{equation}\label{eq-modelselection}
M^k(X) = M^{k-1}(X) \cup \argmin_{\substack{ b\in \mathcal{X}-M^{k-1},\\ a \in M^{k-1}}} \cos^2(a,b),  
\end{equation}

 where $M^0(X)=\{a\}$ and $a$ is the variable with highest $ATE(a)$. We use eq.(\ref{eq-modelselection}) and $0 < K \leq 7$, which selects the bolded variables in Table \ref{table-vars}(first row). We use $K=7$ to match Lalonde's model size. This is a simple variable selection method. It uses only the estimated causal effects and expected dependence among variables. Across-population heterogeneity and others factors readily available in the representation could play roles in more sophisticated criteria.

The method leads to the following selected variables (ordered by the greedy selection),

\begin{equation}
\mathcal{X}=\{ work\_ethics, african\_american, school, worked, drink, nyc\}.
\end{equation}\label{eq-all-covars}
  
The \textit{work\_ethics} variable is related to the survey question 'I'll read a list of things some people feel are important in getting ahead in life. Tell me how important each
one is to you?' The answer follows the scale \textit{\{important, unknown, not important\}} and the variable corresponds to the item 'hard work'. Other items were 'luck', 'education', 'knowing the right people' and 'knowing the community' (the item 'education' also appeared as a top-20 variable, Table \ref{table-vars}). The \textit{african\_american} variable indicates the participant's race, similar to a variable selected by Lalonde. The \textit{school} variable indicates whether the participant was in school within the last 6 months. The \textit{worked} variable indicates whether the participant worked less than 40 hours in the previous 4 weeks (prior to assignment). The \textit{drink} variable indicates the participant's answer to 'do you ever drink beer, wine, gin or other hard liquor?' The \textit{nyc} variable indicates that the participant's NSW site location was New York City. Another location (Philadelphia) and an assignment year (1976) also appear in the top-20 list.
 
Similar to Lalonde, we established a correspondence between the NSW and the PSID for these variables. We ignored \textit{nyc} as the PSID has no public location information. We mapped \textit{hardwork} to the 'earning acts' PSID variable (V2941). It is an aggregate of indicators: '[Family] head seldom or never late for work, head rarely or never fails to go to work when not sick, head has extra jobs, head likes to do difficult or challenging things, etc.' And we mapped \textit{drink} to PSID's annual expenditures on alcoholic beverages variable (V2472) divided by income.  
 
 The estimates for this alternative model specification across the previous two samples are depicted in Fig. 5 (main text). These results confirm some of the reasons Heckman et al. ~\citep{RefWorks:doc:5a51473fe4b0e3e5a635f0b5,RefWorks:doc:5a5145dde4b0eeb35a49d118} put forward to explain the poor performance of matching estimators in Lalonde's NSW subsample: 'locations in different labor markets' appear as an effective factor and that the expansion of the 'limited selected observed variables' can improve methods' accuracy. Results suggest that issues like these can, however, be overcome by observational methods by considering missing causes, non-causes, and how to identify them. They also suggest that SFE can be used to both estimate causal effects and help with model specifications for other estimators.

\subsection{Analysis: Bias-Variance Tradeoff}\label{sect-analysis}

We now motivate the choice to model pairwise individual differences with an alternative analytic argument. Consider an outcome difference predictor $\hat{y}_{ij}$ for individual $i$ (i.e., for outcome differences from others). Most effect estimators consider the least-biased estimate for a \emph{population}. We consider, instead, what would be the least-biased estimate for an \emph{individual}. Repeated samples of ${x_i, y_i}$ can increase the estimator's accuracy. As assumed in Rubin's framework ~\citep{RefWorks:doc:5a514c24e4b0eeb35a49d1bb}, these are rarely available, while observations from other individuals are often abundant. We therefore consider the error incurred by $i$ when using an observation from a second individual $j$. Properties for the following dot-product based estimator are well known ~\citep{ml-book}(2001, p. 50), as well as the relationship to the Gram–Schmidt procedure (the same results can also be derived through product distributions ~\citep{product-distr1}). What distinguishes the following is the formulation of an estimator for outcome differences, $y_{ij}$, as opposed to outcomes, $y_i$. 

For $i$ and an observational pair $(ij)$, the observed outcome difference $y_{ij}$ can only be due to attributes in $x_i$ not present in $x_j$. This leads to a 'counterfactual' estimator for effects at the individual-level. According to the estimator, the observed outcome difference between individuals is due to the effect of attributes that only $i$ has, minus the effect of attributes that only $j$ has, $y_{ij} \sim f(x_i{\ominus}x_j) -  f(x_j{\ominus}x_i)$. 

For an individual $i$, observations from other individuals lead to the effect predictor 

 \begin{equation}\label{eq-pairestimator}
\begin{split}
\hat{y}_{ij} =  \frac{1}{\vert x_j-x_i \vert} \sum_{a\in \mathcal{X}} f_i(a)x_i(a) + \varepsilon_{ij} = \frac{\inner{f_i}{x_i-x_j}}{\vert x_i-x_j \vert} + \varepsilon_{ij},  
\end{split}   
 \end{equation}

where $f_i \in \mathbb{R}^m$ is an individual vector of effects in $y$-scale, $x \in [\texttt{-}1,\texttt{+}1]^m$, $E(\varepsilon)=0$ and $Var(\varepsilon)=0$. Due to the sign convention, the estimator sums the effects of variables in $x_i$ but not in $x_j$, subtracts the effects of variables in $x_j$ but not in $x_i$ and cancels out the effect of variables in both. 

 The estimator's squared error loss can be decomposed in 3 components corresponding to a heterogeneity bias, variance and irreducible error $\varepsilon_{ij}$,  

 \begin{equation}\label{eq-errordecomp}
\begin{split}
Err_{ij} &= E[( y_{ij}- \hat{y}_{ij})^2 ],\\
 &= E\Big[ \Big( y_{ij} - \frac{\inner{f_i}{(x_i-x_j)}}{\selfinner{x_i-x_j}} \Big)^2   \Big],\\
 &= [Heter_{ij}^2 + Var_{ij}^2 + \varepsilon_{ij}^2 ],\\
Heter_{ij} &= E[\hat{y}_{ij}] - y_{ij},\\
&=\Big[\frac{1}{2}\Big( \frac{\inner{f_i}{x_i-x_j}}{\selfinner{x_i-x_j}} - \frac{\inner{f_j}{x_i-x_j}}{\selfinner{x_i-x_j}} \Big)\Big]^2  - y_{ij},\\
&= \Big[\frac{1}{2}\frac{\inner{f_i-f_j}{x_i - x_j}}{\selfinner{x_i-x_j}}\Big]^2 - y_{ij},\\
&= \frac{1}{4}\vert f_i-f_j \vert^2 cos(\theta_{ij})^2 -y_{ij},\\
Var_{ij} &= E[\hat{y}_{ij} - E[ \hat{y}_{ij}]],\\
&= \frac{\varepsilon_{ij}^2}{\vert x_i - x_j \vert^2}.
\end{split}
\end{equation}

where $\theta_{ij}$ is the angle between vectors $f_i$ and $x_i-x_j$. The second term is a squared heterogeneity bias, the $y$-amount by which the estimate differs from the mean using other individuals' effects. The last term is the variance, the expected squared deviation around the estimated mean in $y$-amounts. 

According to this, variance and heterogeneity are related to two distances (norms of position differences) between individuals $i$ and $j$. These are, in turn, related to spaces $X$ and $T_y(X)$. Variance is related to distances in the $[\texttt{-}1,\texttt{+}1]^m$ covariate space, $\vert x_i - x_j \vert$, and heterogeneity to distances over effects, $\vert f_i - f_j \vert$. Larger distances in $X$ correspond to treatments over more variables, decreasing the estimator's variance in eq.(\ref{eq-pairestimator}). Larger distances in $T_y(X)$ correspond to estimates between more heterogeneous individuals, increasing the heterogeneity bias. Decreasing this bias increases the estimate's external validity (for the individual, not the sample population), while decreasing variance increases its internal validity. 

Particularly, eq.(\ref{eq-errordecomp}) suggest that, for a given $\theta_{ij}$, there are two sources of bias: the difference in variable effects among individuals, $f_i(a)-f_j(a)$, and the covariate space dimension, $m$.  For the former, an estimate with minimal $Heter_{ij}$ must have $y_{ij} = \frac{1}{4}\vert f_i-f_j \vert$. This corresponds to the minimized residual illustrated in Fig.\ref{fig-triangle}(c) and implemented by eq.(\ref{eq-sol}). For $m$, decreasing the space to a dimension $d < m$ can increase the variance, $Var_{ij}$, in eq.(\ref{eq-pairestimator}) but decrease $Heter_{ij}$. This motivated the introduction of the treatment size penalty $\phi^{cx}_{ij}$. Both the decision to use only variables that differ among pairs and the proposed optimization procedure can therefore be seen as attempts to reduce individual heterogeneity bias, $Heter_{ij}$. 

Heterogeneity is also decreasing with $\cos^2\theta_{ij}$, which, in turn, reflects statistical correlation. This indicates that heterogeneity is maximal for individuals with highly correlated variables (e.g., sharing many attributes) that observe different effects. This motivated the introduction of the treatment balance penalty $\phi^{bl}_{ij}$, which penalized non-orthogonal pairs, as well as the variable selection criteria in eq.(\ref{eq-modelselection}). Together, these considerations suggest a metric space as representation for a sample population, consisting of a set of orthogonal dimensions with correlated covariates ($\cos^2\theta_{ij} \approx 1$) that are minimally heterogeneous ($y_{ij} \approx \frac{1}{4}\vert f_i-f_j \vert \;$). 

Starting with a single individual $i$ and her individual sample $\{x_i,y_i\}$, we start with maximal internal-validity. As we increase the population scope and consider other individuals' samples $\{x_j,y_j\}$, we can increase estimates' external validity. Learning a representation for individual differences, $\hat{y}_{ij}$, allowed for more accurate (individual) effect estimates while inter-individual effect differences, $Heter_{ij}$, were minimized explicitly. This lead to a space that is 'minimal' but that still reflects observed outcome differences, $y_{ij}$.


\clearpage



\end{document}